\documentclass[paper=a4, fontsize=11pt]{scrartcl} 

\usepackage[T1]{fontenc} 
\usepackage{fourier} 
\usepackage[english]{babel} 
\usepackage{amsmath,amsfonts,amsthm} 
\usepackage[english]{babel} 
\usepackage{amsmath,amsfonts,amsthm} 
\usepackage{epsfig}
\usepackage{pbox}
\usepackage{amssymb}

\usepackage{lipsum} 

\usepackage{sectsty} 
\allsectionsfont{\centering \normalfont\scshape} 

\usepackage{fancyhdr} 
\numberwithin{equation}{section} 
\numberwithin{figure}{section} 
\numberwithin{table}{section} 

\newcommand{\horrule}[1]{\rule{\linewidth}{#1}} 

\title{
\normalfont \normalsize
\horrule{0.5pt} \\[0.8cm] 
\huge \textbf{Incremental Redundancy, \\Fountain Codes  and Advanced Topics} \\ 
\horrule{2pt} \\[0.5cm] 
}

\author{\textit{Suayb S. Arslan} \\ \small{141 Innovation Dr. Irvine, CA 92617} \\ \small{suayb.arslan@quantum.com}} 

\date{\normalsize{``\emph{All men by nature desire knowledge." } \textsl{Aristotle} \\ \normalsize{Version 0.2 -- July 2014}}} 

\begin{document}

\maketitle 

\fbox{\parbox{\textwidth}{\texttt{Revision History:}
\begin{itemize}
\item \textit{Jan. 2014}, \texttt{First version is released.}
\item \textit{Feb. 2014}, \texttt{Uploaded on arXiv.org for public access.}
\item \textit{Jul. 2014}, \texttt{Version 0.2 is released.}
\begin{itemize}
\item \texttt{Table of Contents is added for easy reference. }
\item \texttt{Fixes few grammatical errors, changes the organization and extends the section of systematic fountain code constructions}
\end{itemize}
\end{itemize}}}

\newpage

\tableofcontents

\newpage

\section{\textbf{Abstract}}

\hspace{5mm}\textsl{The idea of writing this technical document dates back to my time in Quantum corporation, when I was studying efficient coding strategies for cloud storage applications. Having had a thorough review of the literature, I have decided to jot down few notes for future reference. Later, these tech notes have turned into this document with the hope to establish a common base ground on which the majority of the relevant research can easily be analyzed and compared. As far as I am concerned, there is no unified approach that outlines and compares most of the published literature about fountain codes in a single and self-contained framework. I believe that this document presents a comprehensive review of the theoretical fundamentals of efficient coding techniques for incremental redundancy with a special emphasis on ``fountain coding" and related applications. Writing this document also helped me have a shorthand reference. Hopefully, It'll be a useful resource  for many other graduate students who might be interested to pursue a research career regarding graph codes, fountain codes in particular and their interesting applications. As for the prerequisites, this document may require some background in information, coding, graph and probability theory, although the relevant essentials shall be reminded to the reader on a periodic basis.}

\hspace{5mm} \textsl{Although various aspects of this topic and many other relevant research are deliberately left out, I still hope that this document shall serve researchers' need well. I have also included several exercises for the warmup. The presentation style is usually informal and the presented material is not necessarily rigorous. There are many spots in the text that are product of my coauthors and myself, although some of which have not been published yet. Last but not least,  I cannot thank enough Quantum Corporation who provided me and my colleagues ``the appropriate playground" to research and leap us forward in knowledge. I cordially welcome any comments, corrections or suggestions.}

\hfill \textsl{January 7, 2014}

\newpage

\section{\textbf{Introduction}}

\hspace{5mm} Although the problem of transferring the information meaningfully is as old as the humankind, the discovery of its underlying mathematical principles dates only fifty years back when Claude E. Shannon introduced the formal description of information in 1948 \cite{Shannon0}. Since then, numerous efforts have been made to achieve the limits set forth by Shannon. In his original description of a typical communication scenario, there are two parties involved; the sender or transmitter of the information and the receiver. In one application, the sender could be writing information on a magnetic medium and the receiver will be reading it out later. In another, the sender could be transmitting the information to the receiver over a physical medium such as twisted wire or air. Either way, the receiver shall receive the corrupted version of what is transmitted. The concept of \emph{Error Correction Coding} is introduced to protect information due to channel errors. For bandwidth efficiency and increased reconstruction capabilities, incremental redundancy schemes have found widespread use in various communication protocols.

\hspace{5mm} In this document, you will be able to find some of the recent developments in ``Incremental redundancy" with a special emphasis on \emph{fountain coding} techniques from the perspective of what was conventional to what is the trend now. This paradigm shift as well as the theoretical/practical aspects of designing and analyzing modern fountain codes shall be discussed. Although there are few introductory papers published in the past such as  \cite{MacKay}, this subject has broadened its influence, and expanded its applications so large in the last decade that it has become impossible to cover all of the details. Hopefully, this document shall cover most of the recent advancements related to the topic and enables the reader to think about ``what is the next step now?" type of questions.

\hspace{5mm}   The document considers fountain codes to be used over erasure channels although the idea is general and  used over error prone wireline/wireless channels with soft input/outputs. Indeed an erasure channel model is more appropriate in a context where fountain codes are frequently used at the application layer with limited access to the physical layer. We also need to note that this version of our document focuses on linear fountain codes although a recent progress has been made in the area of non-linear fountain codes and its applications such as found in Spinal codes \cite{Spinal}. Non-linear class of fountain codes have come with interesting properties due to their construction such as polynomial time \emph{bubble} decoder and generation of coded symbols at one encoding step.

\hspace{5mm} What follows is a set of notation below we use throughout the document and the definition of Binary Erasure Channel (BEC) model.

\subsection{Notation}

\hspace{5mm} Let us introduce the notation we use throughout the document.

\begin{itemize} \small
\item[$\bullet$] Pr\{$A$\} denotes the probability of event $A$ and Pr\{$A$|$B$\} is the conditional probability of event $A$ given event $B$.
\item[$\bullet$] Matrices are denoted by bold capitals ($\textbf{X}$) whereas the vectors are denoted by bold lower case letters ($\textbf{x}$).
\item[$\bullet$] $x_i$ is the $i$-th element of vector $\textbf{x}$ and $\textbf{x}^T$ denotes the transpose of $\textbf{x}$.
\item[$\bullet$] $\mathbb{F}_q$ is a field of $q$ elements. Also $\mathbb{F}_q^k$ is the vector space of dimension $k$ where entries of field elements belong to $\mathbb{F}_q$.
\item[$\bullet$] For $a \in \mathbb{F}_q$, $\textbf{a}$ denotes the all-$a$ vector i.e., $\textbf{a} = [a \ a \ \dots \ a]$.
\item[$\bullet$] $weight(\textbf{x})$ returns the number of nonzero entries of vector $\textbf{x}$.
\item[$\bullet$] $\lfloor . \rfloor$ is the floor and $\lceil . \rceil$ is the ceiling function.
\item[$\bullet$] $f^{\prime}(x)$ and $f^{\prime\prime}(x)$ are the first and second order derivatives of the continuous function $f(x)$. More generally, we let $f^{j}(x)$ to denote $j$-th order derivative of $f(x)$.
\item[$\bullet$]  Let $f(x)$ and $g(x)$ be two functions defined over some real support.  Then, $f(x) = O(g(x))$ if and only if there exists a  $C\in\mathbb{R}^+$ and a real number $x_0\in\mathbb{R}$ such that $|f(x)| \leq C |g(x)|$ for all $x > x_0$.
\item[$\bullet$] coef($f(x),x^j$) is the $j$-th coefficient $f_j$ for a power series $f(x) = \sum_j f_j x^j$.
\item[$\bullet$] For a given set $\mathcal{S}$, $|\mathcal{S}|$ denotes the cardinality of the set $\mathcal{S}$.
\normalsize
\end{itemize}
\hspace{5mm} Since graph codes are essential part of our discussion, some graph theory related terminology might be very helpful.
\begin{itemize} \small
\item[$\bullet$] A graph $G$ consists of the tuple ($V,E$) i.e., the set of vertices (nodes) $V$ and edges $E$.
\item[$\bullet$] A neighbor set (or neighbors) of a node $v$ is the set of vertices adjacent to $v$, i.e., $\{u \in V | u \not= v, (u,v)\in E\}$.
\item[$\bullet$] The degree of a node $v$ is the number of neighbors of $v$.
\item[$\bullet$] A path in graph $G$ is a sequence of nodes in which each pair of consecutive nodes is connected by an edge.
\item[$\bullet$] A cycle is a special path in which the start and the end node is the same node.
\item[$\bullet$] A graph $G$ with $k$ nodes is connected if there is a path between every $\binom{k}{2}$ pair of nodes.
\item[$\bullet$] A connected component of a graph $G$  is a connected subgraph that is not connected to any other node in $G$.
\item[$\bullet$] A giant component of $G$ is a connected component containing a constant fraction of vertices (nodes) of $G$.
\normalsize
\end{itemize}

\begin{figure}[t!]
\centering
\includegraphics[angle=0, height=40mm, width=50mm]{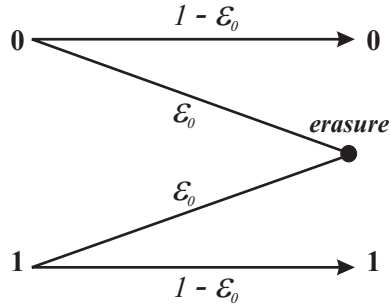}
\caption{Binary Erasure Channel Model.}\label{fig:BEC}
\end{figure}

\subsection{Channel Model and Linear Codes}

\hspace{5mm} Let us consider the channel model we use. The BEC is the simplest non-trivial channel model, yet applicable to real life transmission scenarios. It was introduced by Elias as a toy example in 1954. Indeed erasure channels are usually used to model packet switched networks such as internet, over which erasure codes can be used to reliably transfer information. In a typical transmission scenario, the message files are chopped into $k$ message packets and later are encoded into a set of $n$ packets and transported as independent units through various links. The packet either reaches the destination without any interruption or it is lost permanently i.e., the information is never corrupted. Moreover, the original order of packets may or may not be preserved due to random delays. The destination reconstructs the original $k$ packets if enough number of encoded packets are reliably received.

\hspace{5mm} The BEC channel model is parameterized by the erasure probability $\epsilon_0$ and typically represented as shown in Fig. \ref{fig:BEC}. Let a binary random variable $X \in \{0,1\}$ represent the channel input and is transmitted over the BEC. The output we receive is another random variable $Y \in \{0,1,\mathfrak{e}\}$ where $\mathfrak{e}$ represents an erasure. The mathematical model of this channel is nothing but a set of conditional probabilities given by $Pr\{Y = X |X\} = 1-\epsilon_0$,  $Pr\{Y \not= X |X\} = 0$ and $Pr\{Y = \mathfrak{e} |X\} = \epsilon_0$. Since erasures occur for each channel use independently, this channel is called \emph{memoryless}.  The capacity (the maximum rate of transmission that allows reliable communication) of the BEC is $1-\epsilon_0$ bits per channel use.  A rate $r = k/n$ practical ($n,k$) block code may satisfy $r \leq 1 - \epsilon_0$ while at the same time provide an adequately reliable communication over the BEC. On the other hand, Shannon's so called \emph{channel coding theorem} \cite{Shannon0} states that there is no code with rate $r > 1 - \epsilon_0$ that can provide reliable communication. Optimal codes are the ones that have a rate $r = 1 -\epsilon_0$ and provide zero data reconstruction failure probability. The latter class of codes are called capacity-achieving codes over the BEC.

\hspace{5mm} For an ($n,k$) block code, let $\rho_j$ be the probability of correcting $j$ erasures. Based on one of the basic bounds of coding theory (Singleton bound), we ensure that for $j > n-k$, $\rho_j = 0$. The class of codes that have $p_j = 1$ for $j \leq n-k$ are called Maximum Distance Separable (MDS) codes. This implies that any pattern of $n-k$ erasures can be corrected and hence these codes achieve the capacity of the BEC with erasure probability $(n-k)/n = 1 - k/n$ for large block lengths. The details can be found in any elementary coding book.

 \subsection{Incremental Redundancy}

\begin{figure}[t!]
\centering
\includegraphics[angle=0, height=72mm, width=145mm]{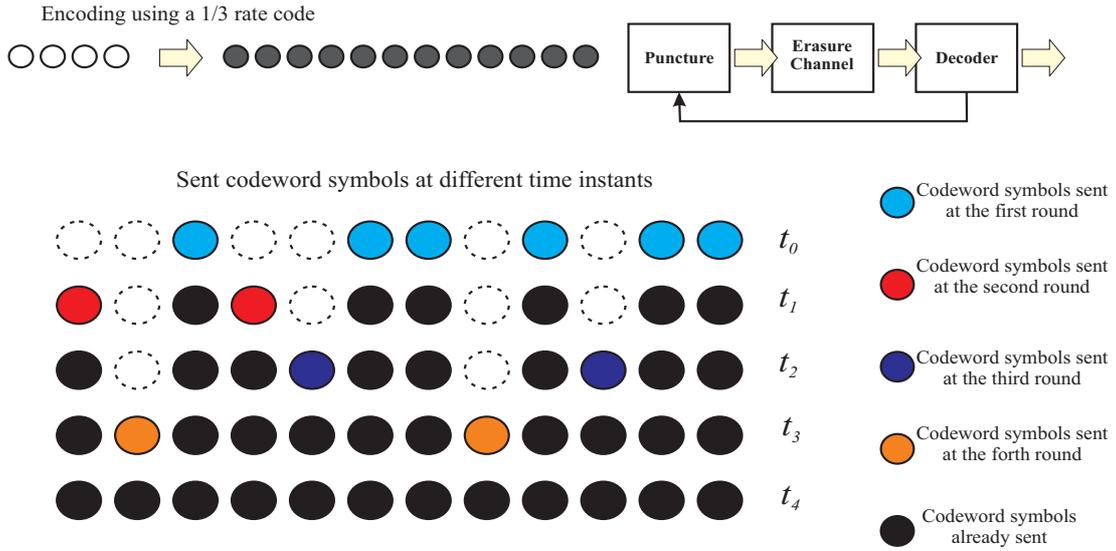}
\caption{An incremental redundancy scheme based on a fixed block code.}\label{fig:IR}
\end{figure}

\hspace{5mm} The idea of \emph{Incremental Redundancy} is not completely new to the
coding community. Conventionally, a low rate block code is punctured i.e., some of the codeword symbols are deliberately deleted
to be able to produce higher rate codes. The puncturing is carefully performed
so that each resultant high rate code performs as equally well as the
same  fixed-rate  block code, although this objective might not be achievable for every class of block codes. In other words, a rate-compatible family of codes has the 
property that codewords of the higher rate codes in the family are prefixes of those of the lower rate ones. Moreover, a perfect family of such codes is the one in which
each element of the family is capacity achieving. For a given channel model, design of a perfect family, if it is ever possible, is of tremendous interest to the research community.

\hspace{5mm} In a typical rate-compatible transmission, punctured symbols
are treated as \emph{erasures} by some form of labeling at the decoder and decoding is initiated afterwards. Once a decoding
failure occurs, the transmitter sends off the punctured symbols one at a time until  the decoding process is successful. If all the
punctured symbols are sent and the decoder is still unable to decode
the information symbols, then a retransmission is initiated through an Automatic Repeat reQuest (ARQ) mechanism.
Apparently, this transmission scenario is by its nature ``rateless" and it provides the desired
incremental redundancy for the reliable transmission of data. An example is shown for a $1/3$-rate block code for encoding a message block
of four symbols in Fig. \ref{fig:IR}. Four message symbols are encoded to produce twelve coded symbols. The encoder sends off six coded symbols (at time $t_0$) through the erasure channel according to a predetermined puncturing pattern. If the decoder is successful, it means a $2/3$-rate code is used (because of puncturing) in the transmission and it transferred message symbols reliably. If the decoder is unable to decode the message symbols perhaps because the channel introduces a lot of erasures, a feedback message is generated and sent.  Upon the reception of the feedback, the encoder is triggered to send two more coded symbols (at time $t_1$) to help the decoding process. This interaction continues until either the decoder sends a success flag or the encoder depletes of any more coded symbols (at time $t_4$) in which case an ARQ mechanism must be initiated for successful decoding. For this example, any generic code with rate $1/3$ should work as long as the puncturing patterns are appropriately designed for the best performance. Our final note is that the rate of the base code (unpunctured) is determined before the transmission and whenever the channel is bad, the code performance may fall apart and ARQ mechanism is inevitable for a reliable transfer.

\hspace{5mm} This rate
compatible approach is well explored in literature and applied to
Reed Solomon codes, Convolutional codes and Turbo codes successfully. Numerous efforts
have been made in literature for developing very good performance Rate Compatible Reed Solomon
(RCRS), Rate Compatible Punctured Convolutional (RCPC) Codes \cite{Hagenauer},
Rate Compatible Turbo Codes (RCTC) \cite{RCPT} and Rate Compatible LDPC codes \cite{RCLDPC} for various applications.
However, the problem with those constructions is that a good set of puncturing
patterns only allows  a limited set of code rates, particularly
for convolutional and turbo codes. Secondly, the decoding process is
almost always complex even if the channel is in good state. It is
because the decoder always decodes the same low rate code. In
addition, puncturing a sub-optimal low rate code may produce
very bad performance high rate code. Thus, the design of the low rate code  as
well as the puncturing table used to generate that code has to be designed very carefully \cite{Oberg}. This usually
complicates the design of the rate compatible block code. In fact, from a purely information theoretic perspective the problem of rateless transmission is well understood and for channels possessing a single maximizing input distribution, a randomly generated linear codes from that distribution will be performing pretty well with high probability. However, construction of such good codes with computationally efficient encoders and decoders is not so straightforward. In order
to save design framework from those impracticalities of developing rate compatible
or in general rateless codes, we need a completely different paradigm  for constructing
codes with rateless properties.

\hspace{5mm} In the next section, you will be introduced to a class of codes that belongs to ``near-perfect code family for erasure channels" called fountain codes. The following chapters
shall explore theoretical principles as well as practical applicability of such codes to real life transmission scenarios. The construction details of such codes have very interesting, deep-rooted relationship to graph theory of mathematics, which will be covered as an advanced topic later in the document.  

\section{\textbf{Linear Fountain Codes}}

\hspace{5mm} Fountain codes (also called modern rateless codes \footnote{
The approach used to transmit such codes is called Digital Fountain
(DF), since the transmitter can be viewed as a fountain emitting
coded symbols until all the interested receivers (the sinks) have
received the number of symbols required for successful decoding.})
are a family of erasure codes where the rate, i.e. the number of
coded and/or transmitted symbols, can be adjusted on the fly. These codes
differ from standard channel codes that are characterized by a rate,
which is essentially selected in the design phase. A fountain encoder can
generate an arbitrary number of coded symbols using simple arithmetic. Fountain codes are primarily introduced for a possible solution to address the information delivery in broadcast and multicast scenarios \cite{Byers}, later though they have found many more fields of application such as data storage. The first known
efficient fountain code design based on this new paradigm is introduced
in \cite{Luby} and goes by the name Luby Transform (LT) codes. LT codes are generated using low density
generator matrices instead of puncturing a low rate code. This low
density generator matrix generates the output encoding symbols to be
sent over the erasure channel.

\hspace{5mm} Without loss of generality, we will focus on the binary
alphabet as the methods can be applied to larger alphabet sizes. An LT
encoder takes a set of $k$ symbols of information to generate coded
symbols of the same alphabet. Let a binary information block
$\textbf{x}^T = (x_1,x_2,\dots,x_k) \in \mathbb{F}^k_2$ consist of
$k$ bits.  The $m$-th coded symbol (check node or symbol) $y_m$ is generated in the
following way: First, the degree of $y_m$, denoted $d_m$, is chosen
according to a suitable degree distribution $\Omega(x) =
\sum_{\ell=1}^k \Omega_\ell x^\ell$ where $\Omega_\ell$ is the
probability of choosing degree $\ell\in \{1,\dots, k \}$. Then,
after choosing the degree $d_m\in \{1,\dots, k \}$, a $d_m$-element
subset of $\textbf{x}$ is chosen randomly according to a suitable selection
distribution. For standard LT coding \cite{Luby}, the selection distribution is the  uniform
distribution. This corresponds to generating a random column vector
$\textbf{w}_m$ of length $k$, and $weight(\textbf{w}_m)=d_m$
positions are selected from a uniform distribution to be logical 1 (or any non-zero element of $\mathbb{F}_q$ for non-binary coding),
without replacement. More specifically, this means that any possible
binary vector of weight $d_m$ is selected with probability
$1/\binom{k}{d_m}$. Finally, the coded symbol is given by $y_m =
\textbf{w}_m^T \textbf{x}$ (mod 2) $m=1,2,\dots N$ \footnote{We
assume that the transmitter sends $\{\textbf{w}_m\}_{m=1}^N$ through
a reliable channel to the decoder. Alternatively,
$\{\textbf{w}_n\}_{n=1}^N$ can be generated pseudo randomly by
initializing it with a predetermined seed at both the encoder and the
decoder. In general, different seeds are used for degree generation and selection of edges after the degrees are determined.}. Note that all these operations are in modulo 2. Some of the coded symbols are erased by the channel, and for decoding
purposes, we concern ourselves only with those $n \leq N$ coded symbols
which arrive unerased at the decoder. Hence the subscript $m$ on
$y_m$, $d_m$ and $\textbf{w}_m$ runs only from 1 to $n$, and we
ignore at the decoder those quantities associated with erased
symbols. If the fountain code is defined over $\mathbb{F}_q^k$, than all the encoding operations follow the rules of the field algebra.

\hspace{5mm} From the previous description, we realize that the encoding process is done by generating binary vectors. The generator matrix of the code is hence
a $k \times n$ binary matrix\footnote{This is convention specific. We could have treated the transpose of \textbf{G} to be the generator matrix as well.} with $\textbf{w}_m$ s as being its column vectors i.e.,
\[ \textbf{G} =  \left( \begin{array}{cccccccccc}
\textbf{w}_1  \ \ | \ \ \textbf{w}_2 \ \ | \ \  \textbf{w}_3   \dots  \textbf{w}_{n-1} \ \ | \ \  \textbf{w}_n \end{array} \right)_{k \times n} \]

\hspace{5mm}   The decodability of the code is in general sense tied to the invertibility of the generator matrix. As will be explored later, this type of decoding is optimal but complex to implement. However, it is useful to draw fundamental limits on the performance and complexity of fountain codes.  Apparently, if $k > n$, the matrix \textbf{G} can not have full rank. If $n \geq k$ and \textbf{G} contains an invertible $k \times k$ submatrix, then we can invert the encoding operation and claim that the decoding is successful.  Let us assume, the degree distribution to be binomial with $p_0=0.5$. In otherwords, we flip a coin for determining each entry of the column vectors of $\textbf{G}$. This idea is quantified for this simple case in the following theorem \cite{Di}.

\hspace{5mm} \textbf{Theorem 1:} \emph{Let us denote the number of binary matrices of dimension $m_x \times m_y$ and rank $R$ by $M(m_x,m_y,R)$. Without loss of generality, we assume $m_y \geq m_x$. Then, the probability that a randomly chosen \textbf{G} has full rank i.e., $R = m_x$ is given by $\prod_{i=0}^{m_x-1}(1-2^{i-m_y})$ }.

\hspace{5mm} PROOF: Note that the number of $1 \times m_y$ matrices with $R=1$ is $2^{m_y}-1$ due to the fact that any non-zero vector has rank one. By induction, the number of ways for extending a $(m_x-1) \times m_y$ binary matrix of rank $(m_x-1)$ to a $m_x \times m_y$ binary matrix of rank $m_x$ is $(2^{m_y} - 2^{m_x-1})$. Therefore, we have the recursion given by
\begin{align}
M(m_x,m_y,m_x) = M(m_x-1,m_y,m_x-1)(2^{m_y} - 2^{m_x-1})
\end{align}

\hspace{5mm} Since $M(1,m_y,1) = 2^{m_y}-1$ and using the recursive relationship above, we will end up with the following probability,
\begin{align}
\frac{M(m_x,m_y,m_x)}{\textrm{All possible binary matrices}}= \frac{M(m_x,m_y,m_x)}{2^{m_xm_y}} = \frac{\prod_{i=0}^{m_x-1}(2^{m_y} - 2^{i})}{2^{m_xm_y}} = \prod_{i=0}^{m_x-1}(1 - 2^{i-m_y})
\end{align}
which proves what is claimed. \hfill $\blacksquare$

\hspace{5mm} For large $m_x = k$ and $m_y = n$, we have the following approximation,
\begin{eqnarray}
\prod_{i=0}^{k-1}(1-2^{i-n}) &\approx& 1 - \left(2^{-n} + 2^{1-n} + \dots + 2^{k-1-n}\right) \\
&=& 1 - 2^{-n}\frac{1-2^{k}}{1-2} = 1 - 2^{-n}(2^{k}-1) \\
&\approx& 1 - 2^{k-n}
\end{eqnarray}

 \hspace{5mm}  Thus, if we let $\gamma = 2^{k-n}$, the probability that $ \textbf{G}$ does not have a full rank and hence is not invertible is given by $\gamma$. We note that this quantity is exponentially related to the extra redundancy $n-k$, needed to achieve a reliable operation. Reading this conclusion reversely, it says that the number of bits required to have a success probability of $1-\gamma$ is given by $n \approx k + \log_2(1/\gamma)$.

 \hspace{5mm}   As can be seen, the expected number of operations for encoding one bit in case of binary fountain codes is the average number of bits XOR-ed to generate a coded bit. Since the entries of $ \textbf{G}$ are selected to be one or zero with half probability, the expected encoding cost per bit is $O(k)$. If we generate $n$ bits, the total expected cost will be $O(nk)$. The decoding cost is the inversion of $ \textbf{G}$ and the multiplication of the inverse with the received word. The cost of the matrix inversion is in general requires an average of $O(n^3)$ operations and the cost of multiplying the inverse is $O(k^2)$ operations. As can be seen the random code so generated performs well with exponentially decaying error probability, yet its encoding and decoding complexity is high especially for long block lengths.

 \hspace{5mm}  A straightforward balls and bins argument might be quite useful to understand the dynamics of edge connections and its relationship to the performance \cite{MacKay}. Suppose we have $k$ bins and $n$ balls to throw into these bins. A throw is performed independent of the successive throws. One can wonder what is the probability that one bin has no balls in it after $n$ balls are thrown. Let us denote this probability by $\omega$. Since balls are thrown without making any distinction between bins, this probability is given by
 \begin{eqnarray}
\omega = \left(1 - \frac{1}{k} \right)^n \approx e^{-n/k}
 \end{eqnarray}
 for large $k$ and $n$. Since the number of empty bins is binomially distributed with parameters $k$ and $p$, the average number of empty bins is $k\omega =  ke^{-n/k}$. In coding theory, a convention for decoder design is to declare a failure probability $\gamma > 0$ beyond which the performance is not allowed to degrade. Adapting such a convention, we can bound the  probability of having at least one bin with no balls $(= \Sigma_k )$ by $\gamma$. Therefore, we guarantee that every single bin is covered with at least one ball with probability greater than $1-\gamma$. We have,
 \begin{eqnarray}
 \Sigma_k = 1 - \left(1 - e^{-n/k} \right)^k < \gamma
 \end{eqnarray}

 For large $n$ and $k$, $\Sigma_k \approx k e^{-n/k} < \gamma$. This implies that $n > k \ln(k/\gamma)$ i.e., the number of balls must be at least scaling with $k$ multiplied by the logarithm of $k$. This result has close connections to the \emph{Coupon collector's problem} of the probability theory, although the draws in the original problem statement have need made with replacement. Finally, we note that this result establishes an information theoretic lower bound on $n$ for fountain codes constructed as described above.

 \hspace{5mm} \textbf{Exercise 1:} Let any length-$k$ sequence of $\mathbb{F}_2^k$ be equally probable to be chosen. The nonzero entries of any element of $\mathbb{F}_2^k$ establishes the indexes of message symbols that contribute to the generated coded symbol. What is the degree distribution in this case i.e., $\Omega(x)$? Is there a closed form expression for $\Omega(x)$?

  \hspace{5mm} \textbf{Exercise 2:} Extend the result of Exercise 1 for $\mathbb{F}_q^k$.

\subsection{High density Fountain codes:  Maximum Likelihood Decoding}

\hspace{5mm} The  ML decoding over the BEC is the problem of recovering $k$
information symbols from the $n$ reliably received  coded (check) symbols i.e., solving a system of $n$ linear equations for $k$ unknowns. As theorem 1 conveys this pretty nicely, the decoder failure probability (message block failure) is the  probability that $\textbf{G}^T$ has not rank $k$. We can in fact extend the result of theorem 1 to $q-$ary constructions (random linear codes defined over $\mathbb{F}_q$) using the same argument to obtain the probability that $\textbf{G}^T$ has rank $k$, given by
\begin{eqnarray}
\prod_{i=0}^{m_x-1}(1 - q^{i-m_y}) = (q^{-m_y};q)_{m_x}
\end{eqnarray}
where the entries of $\textbf{G}^T$ are selected uniform randomly from $\mathbb{F}_q$ and $(a;q)_{k} \triangleq \prod_{i=0}^{k-1}(1-aq^i)$ is $q-$ Pochhammer symbol. The following theorem shall be useful to efficiently compute the failure probability of dense random fountain codes.

\hspace{5mm} \textbf{Theorem 2:} \emph{The ML decoding failure probability of dense random linear fountain code with the generator matrix \textbf{G} defined over $\mathbb{F}_q$ satisfies the following inequalities
\begin{eqnarray}
q^{m_x-m_y-1} \leq 1- \prod_{i=0}^{m_x-1}(1 - q^{i-m_y}) < \frac{q^{m_x - m_y}}{q-1}
\end{eqnarray}}

\hspace{5mm} PROOF: The lower bound follows from the observation that $1 - q^{m_x-1-m_y} \geq  \prod_{i=0}^{m_x-1}(1 - q^{i-m_y})$ due to each term in the product is $\leq 1$. The upper bound can either be proved by induction similar to theorem 1, as given in \cite{Liva} or using union bound arguments as in theorem 3.1 of \cite{Amin3}. Here what is interesting is the upper as well as the lower bounds are independent of $m_x$ or $m_y$ but depends only on the difference $m_y - m_x$.   \hfill $\blacksquare$

\begin{figure}[t!]
\centering
\includegraphics[angle=0, height=70mm, width=100mm]{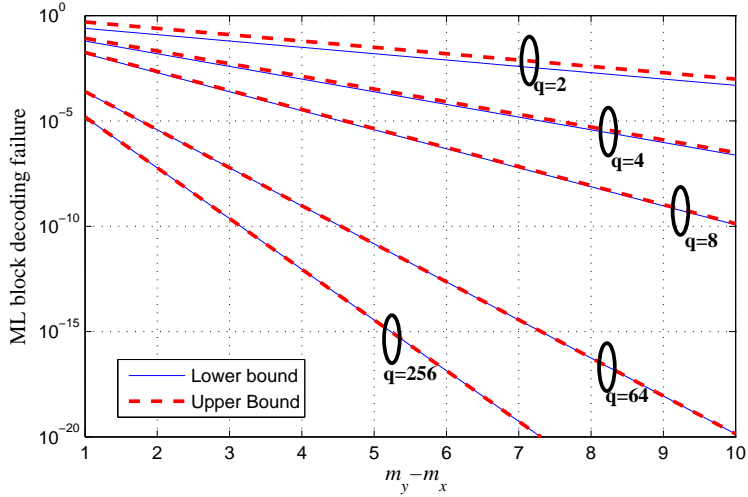}
\caption{Lower and upper bounds on the  block ML decoding failure of dense random linear fountain codes, defined over $\mathbb{F}_q$ for $q=2,4,8,64, 256$. }\label{fig:MLq}
\end{figure}

\hspace{5mm} The bounds of theorem 2 are depicted in Fig. \ref{fig:MLq} as functions of $m_y - m_x$, i.e., the extra redundancy. As can be seen bounds converge for large $q$ and significant gains can be obtained over the dense fountain codes defined over $\mathbb{F}_2$. These performance curves can in a way be thought as the lower bounds for the rest of the fountain code discussion of this document. In fact, here we characterize what is achievable without thinking about the complexity of the implementation. Later, we shall mainly focus on low complexity alternatives while targeting a performance profile close to what is achievable.

\hspace{5mm} The arguments above and of previous section raises curiosity about the symbol-level ML decoding  performance of dense random fountain codes i.e., a randomly generated dense $\textbf{G}^T$ matrix of size $n \times k$. Although exact formulations might be cumbersome to find, tight upper and lower bounds have been developed in the past. Let us consider it over the binary field for the moment and let $p_0$ be the probability of selecting any entry of $\textbf{G}^T$  to be one. Such an assumption induces a probability distribution on the check node degrees of the fountain code. In fact it yields the following degree distribution,
\begin{eqnarray}
\sum_{d=0}^{k} \Omega_d x^d = \sum_{d=0}^{k} \binom{k}{d} p_0^d (1-p_0)^{k-d} x^d
\end{eqnarray}

\hspace{5mm} If we assume $\Omega_0 = 0$, then we need to normalize the binomial distribution so that $\Omega(1) = 1$. Our previous argument for balls/bins establishes a lower bound on any decoding algorithm used for fountain codes. This is because if an input node is not recovered by any one of the coded/check symbols, no decoding algorithm can recover the value associated with that input node.

\hspace{5mm} Let $v$ be any message symbol and condition on the degree $d$ of a coded symbol, the probability of that $v$ is not used in the computation of that coded symbol is given by $1 - d/k$. The unconditional probability shall be given by,
\begin{eqnarray}
\sum_d \Omega_d \left(1 - \frac{d}{k}\right) = 1 - \frac{\Omega^{\prime}(1)}{k}
\end{eqnarray}

\hspace{5mm} If $n$ independently generated coded symbols are collected for decoding, the probability that none of them are generated using the value of message symbol $v$ shall be bounded above and below by
\begin{eqnarray}
e^{-\frac{\Omega^{\prime}(1)n}{k-\Omega^{\prime}(1)}} \leq \left( 1 - \frac{\Omega^{\prime}(1)}{k} \right)^n \leq e^{-\frac{\Omega^{\prime}(1)n}{k}} \label{fig:LB1}
\end{eqnarray}

 \hspace{5mm} \textbf{Exercise 3:} Show the lower bound of equation (\ref{fig:LB1}). Hint: You might want to consider the Taylor series expansion of $\ln(1-x)$.

 \hspace{5mm} If we let $0 < h_0 < 1$ be the denseness parameter such that $\Omega^{\prime}(1) = h_0k$, the lower bound will be of the form $(1 - h_0)^n \approx 1 - h_0k$ for $h_0 << 1$. In otherwords, larger $h_0$ means sharper fall off (sharper slope) i.e., improved lower bound. This might give us a hint that denser generator matrices shall work very well under optimal (ML) decoding assumption. We will explore next an upper bound on the symbol level performance of ML decoding for  fountain codes. It is ensured by our previous argument that equation (\ref{fig:LB1}) is a lower bound for the ML decoding. Following theorem from \cite{Rahnavard} establishes an upper bound for ML decoding,

\hspace{5mm} \textbf{Theorem 3:} \emph{For a fountain code of length $k$ with a degree distribution $\Omega(x)$ and collected number of coded symbols $n$, symbol level ML decoding performance can be upper bounded by
\begin{eqnarray}
\sum_{l = 1}^k \binom{k-1}{l-1} \left[\sum_d \Omega_d \frac{\sum_{s = 0,2,\dots, \min\{l, \lfloor d \rfloor_{even} \} } \binom{l}{s}\binom{k-l}{d-s}}{\binom{k}{d}}\right]^n
\end{eqnarray}
where $\lfloor.\rfloor_{even}$ rounds down to the nearest even integer.}

\hspace{5mm} PROOF: Probability of ML decoding  failure $P_{ML}(b)$ can be thought as the probability that any arbitrary $i$-th bit ($i\in \{1,2,\dots,k\}$) cannot be recovered.
\begin{align}
P_{ML}(b) =& Pr\left\{\exists x \in \mathbb{F}_2^{k}, x_i = 1 \ni \textbf{G}^T\textbf{x} = \textbf{0}\right\}  \label{Union1} \\
\leq& \sum_{\textbf{x} \in \mathbb{F}_2^{k},  x_i = 1}  Pr\left\{  \textbf{G}^T\textbf{x} = \textbf{0} \right\} \label{Union2}
\end{align}
where $\textbf{G}^T\textbf{x} = \textbf{0}$ indicates that $i$-th row of $\textbf{G}$ is dependent on a subset of rows of  $\textbf{G}$ and hence causing the rank to be less than $k$. We go from (\ref{Union1}) to  (\ref{Union2}) using the union bound of events. Note that each column of $\textbf{G}$ is independently generated i.e., different realizations of a vector random variable $\textbf{w} \in F_2^k$. Therefore, we can write
\begin{eqnarray}
Pr\left\{  \textbf{G}^T\textbf{x} = \textbf{0} \right\} = \left( Pr\left\{  \textbf{w}^T\textbf{x} = 0 \right\} \right)^n \label{fig:Omegax}
\end{eqnarray}

\hspace{5mm} Let us define $\textbf{v} = \{w_1x_1, w_2x_2, \dots, x_kx_k\}$. It is easy to see that the following claim is true.
\begin{eqnarray}
 \textbf{w}^T\textbf{x} = 0 \Leftrightarrow \textrm{\emph{weight}(\textbf{v}) is even}
\end{eqnarray}

Let us condition on \emph{weight}(\textbf{w})=$d$ and \emph{weight}(\textbf{x})=$l$, we have
\begin{eqnarray}
&& Pr\left\{  \textbf{w}^T\textbf{x}  | \emph{weight}(\textbf{w})=d,  \emph{weight}(\textbf{x})=l \right\}   \\
&& \ \ \ \ \ \ \ \ \ \ \ \ \ \ \ \ \ \ \ \ \ \ \ \ \ = Pr\left\{ \textrm{\emph{weight}(\textbf{v}) is even} | \emph{weight}(\textbf{w})=d,  \emph{weight}(\textbf{x})=l \right\} \\
&& \ \ \ \ \ \ \ \ \ \ \ \ \ \ \ \ \ \ \ \ \ \ \ \ \ = \frac{\sum_{s = 0,2,\dots, \min\{l, \lfloor d \rfloor_{even} \} } \binom{l}{s}\binom{k-l}{d-s}}{\binom{k}{d}}
\end{eqnarray}

\hspace{5mm} If we average over $\Omega(x)$ and all possible choices of $\textbf{x}$ with $x_i=1$ and \emph{weight}(\textbf{x})=$l$, using equation (\ref{fig:Omegax}) we obtain the desired result. \hfill $\blacksquare$

\hspace{5mm} In delay sensitive transmission scenarios or storage applications (such as email transactions), the use of short block length codes is inevitable. ML decoding might be the only viable choice in those cases for a reasonable performance. The results of theorem 3 is extended in \cite{Schot} to $q$-ary random linear fountain codes and it is demonstrated that these codes show excellent performance under ML decoding. The tradeoff between the density of the generator matrix (also the complexity of decoding) and the associated error floor is investigated. Although the complexity of ML decoding might be tolerable for short block lengths, it is
computationally prohibitive for long block lengths ($k$ large). In order to allow low decoding complexity with increasing block
length, an iterative scheme called Belief Propagation (BP) algorithm is used.

\subsection{Low Density Fountain codes: BP decoding and LT Codes}

\hspace{5mm}
Although ML decoding is optimal, BP is a more popular algorithm and it generally results in successful decoding as long as the
generator/parity check matrices of the corresponding block codes are
sparsely designed. Otherwise for dense parity check matrices, BP algorithm terminates early and becomes
useless. Let us remember our original example when we mentioned  random  binary fountain codes where each entry of $\textbf{G}$ is one (``non-zero" for binary codes) with probability $p_0 = 0.5$. The probability of a column of $\textbf{G}$ has a degree-one weight is given by $\binom{k}{1} (1/2)^1(1-1/2)^{k-1} = k/2^k$ and as $k \rightarrow \infty$, this probability goes to zero, which makes the BP not a suitable decoding algorithm.

\begin{figure}[t!]
\centering
\includegraphics[angle=0, height=35mm, width=130mm]{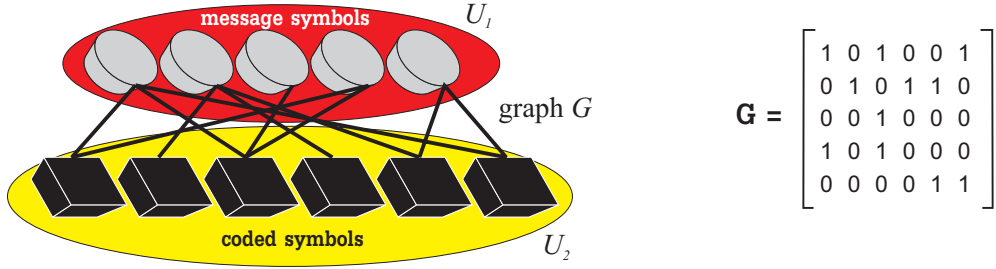}
\caption{Bipartite graph representation of a block code.}\label{fig:BP}
\end{figure}

\hspace{5mm}  In order to describe the BP algorithm, we will use a simple graphical representation of a fountain code using a \emph{bipartite} graph. A bipartite graph is graph in which the set of vertices $V$ can be divided into two disjoint non-empty sets $U_1$ and $U_2$ such that edges in $E$ connects a node in $U_1$ to another node in $U_2$. In fact, any block code can be represented using a bipartite graph, however this representation is particularly important if the code has sparse generator or parity check matrices. Let us consider Fig. \ref{fig:BP}. As can be seen, a coded symbol is generated by adding (XOR-ing for binary domain) a subset of the message symbols. These subsets are indicated by drawing edges between the coded symbols (check nodes) and the message symbols (variable nodes) making up all together a bipartite graph named \emph{G}. The corresponding generator matrix is also shown for a code defined over $\mathbb{F}_2^6$. The BP algorithm has the prior information about the graph  connections but not about the message symbols. A conventional way to transmit the graph connections to the receiver side is by way of a pseudo random number generators fed with one or more seed numbers. The communication is reduced to communicating the seed number which can be transferred easily and reliably. In the rest of our discussions, we will assume this seed is reliably communicated.

\hspace{5mm} BP algorithm can be summarized as follows,
\\ ---------------------------------
\begin{itemize} \small
\item \textbf{Step 1}: Find a coded symbol of degree-one. Decode the unique message symbol that is connected to this coded symbol. Next, remove the edge from the graph.  If there is no degree-one coded symbol, the decoder cannot iterate further and reports a failure.
\item \textbf{Step 2}: Update the neighbors of the decoded message symbol based on the decoded value i.e., each neighbor of the decoded message symbol is added (XOR-ed in binary case) the decoded value. After this update, remove all the neighbors and end up with a reduced graph.
\item \textbf{Step 3}: If there are unrecovered message symbols, continue with the first step based on the reduced graph. Else, the decoder stops.
\end{itemize}
\normalsize
---------------------------------

\begin{figure}[t!]
\centering
\includegraphics[angle=0, height=105mm, width=103mm]{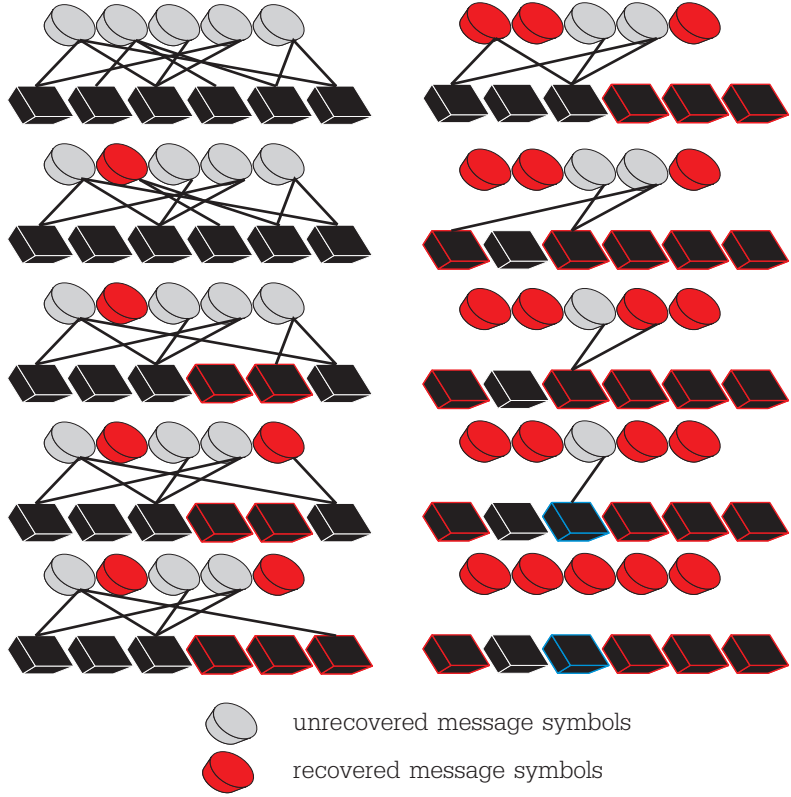}
\caption{A BP decoding algorithm example.}\label{fig:BP2Ex}
\end{figure}

\hspace{5mm}  As is clear from the description of the decoding algorithm, the number of operations is related to the number of edges of the bipartite graph representation of the code, which in turn is related to the degree distribution. Therefore, the design of the fountain code must ensure a good degree distribution that allows low complexity (sparse $\textbf{G}$ matrix) and low failure probability at the same time. Although these two goals might be conflicting, the tradeoff can be solved for a given application.

\hspace{5mm} It is best to describe the BP algorithm through an example. In Fig. \ref{fig:BP2Ex}, an example for decoding the information symbols as well as edge removals (this is alternatively called \emph{graph pruning} process) are shown in detail. As can be seen in this example, in each decoding step only one information symbol is decoded. In fact for BP algorithm to continue, we must have at least one degree-one coded symbol in each decoding step. This implies, the least amount of message symbols to be decoded in each iteration is one if BP algorithm is successful. The set of degree-one message symbols in each iteration is conventionally named as the  \emph{ripple} \cite{Luby}.

\hspace{5mm} Luby proposed an optimal degree distribution $\Omega(x)$ (optimal in expectation) called Soliton Distribution. In Soliton distribution, the expected ripple size is one i.e., the optimal ripple size  for BP algorithm to continue\footnote{Here, we call it optimal because although having ripple size greater than one is sufficient for BP algorithm to continue iterations, the number of extra coded symbols needed to decode all of the message symbols generally increase if the ripple size is greater than one.}. Since it is only expected value, in reality it might be very likely that the ripple size is zero at any point in time. Apparently, degree distributions that show good performance in practice are needed.

\hspace{5mm} In order to derive the so called Soliton distribution, let us start with a useful lemma.

\hspace{5mm} \textbf{Lemma 1:} \emph{Let $\Omega(x) = \sum_d \Omega_d x^d$ be a generic degree distribution and $f(x)$ is some monotone increasing function of  the discrete variable $x > 0$ i.e. $f(x-1) < f(x)$ for all $x$. Then,
\begin{eqnarray}
\Omega^{\prime}(f(x)) -  \Omega^{\prime}(f(x-1)) \leq (f(x) - f(x-1))\Omega^{\prime\prime}(f(x))
\end{eqnarray}
}
\hspace{5mm} PROOF:  Let us consider the first order derivative and write,
\begin{eqnarray}
\Omega^{\prime}(f(x)) -  \Omega^{\prime}(f(x-1)) &=& \sum_d d \Omega_d \left( f(x)^{d-1} - f(x-1)^{d-1}\right) \\
&=&  \sum_d d \Omega_d \left( f(x) - f(x-1)\right) \sum_{j=0}^{d-2} f(x)^{d-j-2} f(x-1)^{j} \\
&=&\left( f(x) - f(x-1)\right) \sum_d d \Omega_d f(x)^{d-2}\sum_{j=0}^{d-2}\left(\frac{f(x-1)}{f(x)}\right)^j \\
&<& \left( f(x) - f(x-1)\right) \sum_d d (d-1) \Omega_d f(x)^{d-2} \\
&=& \left( f(x) - f(x-1)\right) \Omega^{\prime\prime}(f(x))
\end{eqnarray}
which proves the inequality. The equality will only hold asymptotically as will be explained shortly. Note here that since by assumption $f(x-1) < f(x)$ for all $x$, we have $\sum_{j=0}^{d-2}\left(\frac{f(x-1)}{f(x)}\right)^j < d-1$ used to establish the inequality above. \hfill  $\blacksquare$

\hspace{5mm} If we let $f(x) = x / k$ for $k \rightarrow \infty$ and $\Delta x = f(x) - f(x-1) = 1/k$ we will have
\begin{eqnarray}
\lim_{k \rightarrow \infty} \frac{\Omega^{\prime}(f(x)) -  \Omega^{\prime}(f(x-1))}{f(x) - f(x-1)} = \lim_{\Delta x \rightarrow 0} \frac{\Omega^{\prime}(f(x)) -  \Omega^{\prime}(f(x) - \Delta x)}{\Delta x} = \Omega^{\prime\prime}(f(x))
\end{eqnarray}
which along with Lemma 1 shows that there are cases the equality will hold, particularly for asymptotical considerations. In this case therefore, we have for $k \rightarrow \infty$
\begin{eqnarray}
\Omega^{\prime}(x/k) -  \Omega^{\prime}((x-1)/k) \Rightarrow \frac{1}{k} \Omega^{\prime\prime}(x/k) \label{3.26}
\end{eqnarray}

\hspace{5mm} This expression shall be useful in the following discussion.

\begin{figure}[t!]
\centering
\includegraphics[angle=0, height=35mm, width=95mm]{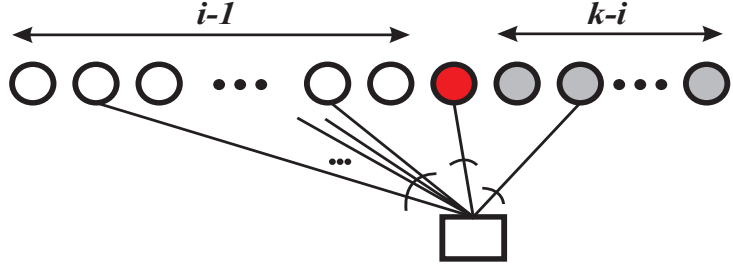}
\caption{A degree release at the $i$-th iteration of a coded symbol of degree $d$.}\label{fig:BP2}
\end{figure}

\hspace{5mm} In the process of graph pruning, when the last edge is removed from a coded symbol, that coded symbol is said to be \emph{released} from the rest of decoding process and no longer used by the BP algorithm. We would like to find the probability that a coded symbol of initial degree $d$ is released at the $i$-th iteration. For simplicity, let us assume edge connections are performed with replacement (i.e., easier to analyze) although in the original LT process, edge selections are performed without replacement. The main reason for this assumption is that in the limit ($k \rightarrow \infty$) both assumptions probabilistically converge to one another. In order to release a degree-$d$ symbol, it has to have exactly one edge connected to $k-i$ unrecovered symbols after the iteration $i$, and not all the remaining $d-1$ edges are connected to $i-1$ already recovered message symbols. This probability is simply given by
\begin{eqnarray}
\binom{d}{1} \left(\frac{k-i}{k}\right)\left( \left(\frac{i}{k}\right)^{d-1} - \left(\frac{i-1}{k}\right)^{d-1} \right)
\end{eqnarray}
in which a message symbol can be chosen in $d$ different ways from $k-i$ message symbols. This is illustrated in Fig. \ref{fig:BP2}. Since we need to have at least one connection with the ``red" message symbol of Fig. \ref{fig:BP2}, we subtract the probability that all the remaining edges make connection with the already recovered $i-1$ message symbols. If we average over the degree distribution we obtain the probability of releasing a coded symbol at iteration $i$, $P_k(i)$ as follows,
\begin{eqnarray}
P_k(i) = \sum_d \Omega_d d \left(\frac{k-i}{k}\right)\left( \left(\frac{i}{k}\right)^{d-1} - \left(\frac{i-1}{k}\right)^{d-1} \right) = \left(1-i/k\right) \left( \Omega^{\prime}(i/k) - \Omega^{\prime}((i-1)/k) \right)
\end{eqnarray}

\hspace{5mm} Asymptotically, we need to collect only $k$ coded symbols to reconstruct the message block. The expected number of coded symbols the algorithm releases at iteration $i$ is therefore $k$ times $P_k$ and is given by (using the result of Lemma 1 and equation (\ref{3.26}))
\begin{eqnarray}
k P_k = k\left(1-i/k\right) \frac{1}{k}\Omega^{\prime\prime}(i/k)  &=& \left(1-i/k\right)\Omega^{\prime\prime}(i/k)
\end{eqnarray}
which must be equal to 1 because at each iteration ideally one and only one coded symbol is released and at the and of $k$ iterations all message symbols are decoded successfully.
If we set $y = i/k$, for $0 < y < 1$, we will have the following differential equation to solve
\begin{eqnarray}
(1-y) \Omega^{\prime\prime}(y) = 1 \label{KeyEqnOmega}
\end{eqnarray}

\hspace{5mm} The general solution to this second order ordinary differential equation is given by
\begin{eqnarray}
\Omega(y) = (1-y)\ln(1-y) + c_1y +c_2
\end{eqnarray}
with the initial conditions $\Omega(0) = 0$ and $\Omega(1) = 1$ (due to sum of probabilities must add up to unity) to find $c_1 = 1 $ and $c_2 = 0$. Using the series expansion for $\ln (1-y) = -\sum_{d=1}^\infty \frac{y^d}{d}$, we obtain the sum i.e., the \emph{limiting distribution},
\begin{eqnarray}
\Omega(y) = (1-y)\ln(1-y) + y &=& - \sum_{d=1}^{\infty} \frac{y^d}{d} + \sum_{n=1}^{\infty} \frac{y^{d+1}}{d} + y \\
&=& - \sum_{d=2}^{\infty} \frac{y^d}{d} + \sum_{n=2}^{\infty} \frac{y^{d}}{d-1} \\
&=& \sum_{d=2}^{\infty} \frac{y^d}{d(d-1)} = \sum_{d} \Omega_d y^d
\end{eqnarray}
from which we see in the limiting distribution $\Omega_1 = 0$ and therefore, the BP algorithm cannot start decoding. A finite length analysis (assuming selection of edges without replacement) show that \cite{Luby} the distribution can be derived to be of the form
\begin{eqnarray}
\Omega(y) = \frac{y}{k} + \sum_{d=2}^{k} \frac{y^d}{d(d-1)} \label{SD1}
\end{eqnarray}
which is named as \emph{Soliton distribution} due to its resemblance to physical phenomenon known as Soliton waves. We note that Soliton distribution is almost exactly the same as the limiting distribution of our analysis. This demonstrates that for $k \rightarrow \infty$, Soliton distribution converges to the limiting distribution.

\hspace{5mm} \textbf{Lemma 2:} \emph{Let $\Omega(x)$ be a Soliton degree distribution. The average degree of a coded symbol is given by $O(\ln (k))$
}.

\hspace{5mm} PROOF:  The average degree per coded symbol is given by
\begin{eqnarray}
\Omega^{\prime}(1) = \sum_{d=1}^k d \Omega_d = 1/k + \sum_{d=2}^k 1/(d-1) = 1/k + \sum_{d=1}^{k-1} 1/d = \sum_{d=1}^{k} 1/d = \ln (k) + 0.57721 + \pi_k
\end{eqnarray}
where 0.57721 is Euler's constant and $\pi_k \rightarrow 0$ as $k \rightarrow \infty$. Thus, $\Omega^{\prime}(1) = O(\ln (k))$ \hfill  $\blacksquare$

\hspace{5mm} The average total number of edges is therefore $O(k \ln (k))$. This reminds us our information theoretic lower bound on the number of edge connections for vanishing error probability. Although Solition distribution achieves that bound, it is easy to see that the complexity of the decoding operation is not linear in $k$ (i.e., not optimal).

\hspace{5mm} Note that our analysis was based on the assumption that the average number of released coded symbols is one in each iteration of the BP algorithm. Although this might be the case with high probability in the limit, in practice it is very likely that in a given iteration of the algorithm, we may not have any degree-one coded symbol to proceed decoding. Although the ideal Soliton distribution works poorly in practice, it gives some insight into somewhat a more robust distribution. The \emph{robust Soliton distribution}
is proposed \cite{Luby} to solve the tradeoff that the expected size of the ripple should be large enough
at each iteration so that the ripple never disappears completely with high probability, while ensuring that the number of coded symbols for decoding the whole message block is minimal.

\hspace{5mm} Let $\gamma$ be the allowed decoder failure probability and Luby designed the distribution (based on a simple random walker argument) so that the ripple size is about $\ln (k/\gamma) \sqrt{k}> 1$ at each iteration of the algorithm. The robust Soliton distribution is given by

\hspace{5mm} \textbf{Definition 1:} \emph{Robust Soliton} distribution $\mu(x)$ (RSD). Let $\Omega(x)$ be a Soliton distribution and $R =
c\ln(k/\gamma)\sqrt{k}$ for some suitable constant $c > 0$ and the allowed decoder failure probability $\gamma > 0$,
\begin{enumerate}
\item[$\bullet$] For
$i=1,\dots,k$: probability of choosing degree $i$ is given by $\mu_i
= \left(\Omega_i + \tau_i\right)/\beta$, where
\begin{enumerate}
\item[$\circ$] $\tau_i =
\begin{cases} R/ik & \text{for $i=1,\dots,k/R-1$,}
\\
R\ln(R/\delta)/k & \text{for $i=k/R$,}
\\
0 & \text{for $i=k/R+1,\dots,k$}
\end{cases}$
\item[$\circ$] $\beta = \sum_{i=1}^k \Omega_i + \tau_i$.
\end{enumerate}
\end{enumerate}

\hspace{5mm} In this description, the average number of degree-$i$ coded symbols is set to $k\left(\Omega_i + \tau_i\right)$. Thus, the average number of coded symbols are given by
\begin{eqnarray}
n = k \left( \sum_{d=1}^k \Omega_d + \tau_d \right) &=& k + \sum_{d=1}^{k/R-1} \frac{R}{i} + R \ln(R/\gamma) \\
&\approx& k + R \ln (k/R-1)  + R \ln(R/\gamma) \\
&\leq& k + R\ln (k/R)\ln(R/\gamma) = k + R\ln (k/\gamma) \\
&=& k + c \ln^2(k/\gamma)\sqrt{k}
\end{eqnarray}

\hspace{5mm} \textbf{Exercise 4:} Show that the average degree of a check symbol using RSD is $\mu^{\prime}(1) = O(\ln(k/\gamma))$. Hint: $\mu^{\prime}(1) \leq \ln(k) + 2 + \ln (R/\gamma)$.

\hspace{5mm}  Exercise 4 shows that considering the practical scenarios that RSD is expected to perform better than the Soliton distribution, yet the average number of edges (computations) are still on the order of $k\ln(k/\gamma)$ i.e., not linear in $k$. Based on our previous argument of lemma 2 and the information theoretic lower bound, we conclude that there is no way (no degree distribution) to make encoding/decoding perform linear in $k$ if we impose the constraint of negligible amount of decoder failure. Instead, as will be shown, degree distributions with a constant maximum degree allow linear time operation. However, degree distributions that have a constant maximum degree results in a decoding error floor due to the fact that with this degree distribution only a fraction of message symbols can be decoded with vanishing error probability. This led research community to the idea of concatenating LT codes with linear time encodable erasure codes to be able to execute the overall operation in linear time. This will be explored in Section 4.

\subsection{Asymptotical performance of LT codes under BP Decoding}

\begin{figure}[t!]
\centering
\includegraphics[angle=0, height=55mm, width=105mm]{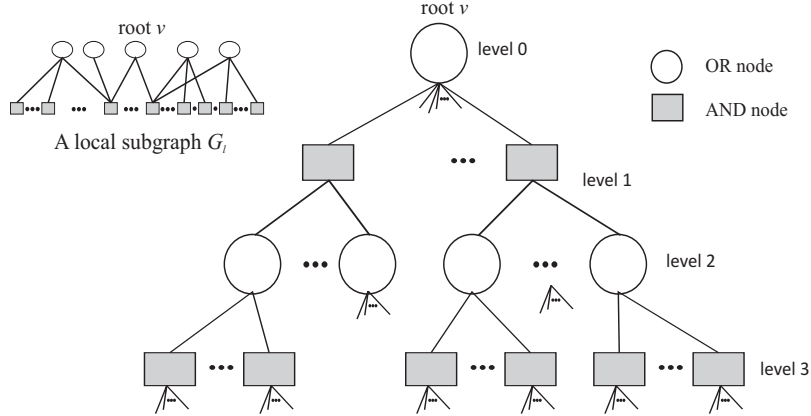}
\caption{A subgraph $G_l$ of the bipartite graph representation of an LT code asymptotically approaches to a tree, which validates the assumption under which the And-Or tree lemma has been derived.}\label{LTtree}
\end{figure}

\hspace{5mm}  Before presenting more modern fountain codes, let us introduce a nice tool for predicting the limiting behavior of sparse fountain codes under BP decoding. We start by constructing a subgraph $G_l$ of $G$ which is the bipartite graph representation of the LT code. We pick an edge connecting some variable node $v$ with the check node $c$ of $G$ uniform randomly. We call $v$ to be the root of $G_l$. Then, $G_l$ will consist of all vertices $V_v$ of $G$ that can be reached from $2l$ hops from $v$ down, by traversing all the edges of $G$ that connects any two vertices of $V_v$. As $n \rightarrow \infty$, it can be shown using Martingale arguments that the resulting subgraph $G_l$ converges to a tree \cite{Petar}. This is depicted in Fig. \ref{LTtree}.

\hspace{5mm} Next, we assume that $G_l$ is a randomly generated tree of maximum depth (level) $2l$ where the root of the tree $v$ is at level $0$. We label the variable (message) nodes at depths $0,2,\dots,2l-2$ to be $OR$-nodes and the check (coded) symbol nodes at depths $1,3,\dots,2l-1$  to be $AND$-nodes. The reason for this assignment is obvious because in the BP setting, variable nodes only send ``one" i.e., they can be decoded if any one of the check symbols send ``one". Similarly, check nodes send ``one" if and only if all of the other adjacent variable nodes send ``one" i.e., that check node can decode the particular variable node if all of the other variable nodes are already decoded and known.

\hspace{5mm} Suppose $OR$ nodes select $i$ children to have with probability $\alpha_i$ whereas $AND$ nodes select $i$ children to have with probability $\beta_i$. These induce two probability distributions $\alpha(x) = \sum_{i=0}^{L_\alpha}\alpha_i x^i$ and $\beta(x) = \sum_{i=0}^{L_\beta}\beta_i x^i$ associated with the tree $G_l$, where $L_\alpha$ and $L_\beta$ are the constant maximum degrees of these probability distributions. Note that the nodes at depth $2l$ are leaf nodes and do not have any children.

\hspace{5mm} The process starts with by assigning the leaf nodes a ``0" or a ``1" independently with probabilities $y_0$ or $y_1$, respectively. We think of the tree $G_l$ as a boolean circuit consisting of $OR$ and $AND$ nodes which may be independently short circuited with probabilities $a$ and $b$, respectively. $OR$ nodes without children are assumed to be set to ``0" and $AND$ nodes without children are assumed to be set to ``1". We are interested in the probability of the root node being
evaluated to 0 at the end of the process. This is characterized by the following popular Lemma \cite{Luby2}.

\hspace{5mm} \textbf{Lemma 3:} \emph{(The And-Or tree lemma) The probability $y_l$ that the root of $G_l$ evaluates to 0 is $y_l = f(y_{l-1})$, where $y_{l-1}$ is the probability that the root node of a $G_{l-1}$ evaluates to 0, and
\begin{eqnarray}
f(x) = (1-a)\alpha(1 - (1-b)\beta(1-x)), \ \ \ \alpha(x) = \sum_{i=0}^{L_\alpha}\alpha_i x^i \ \ \  \textrm{and} \ \ \ \beta(x) = \sum_{i=0}^{L_\beta}\beta_i x^i
\end{eqnarray}
}

\hspace{5mm} PROOF:  Proof is relatively straightforward as can be found in many class notes given for LDPC codes used over BECs. Please also see \cite{Petar} and \cite{Luby2}. \hfill  $\blacksquare$

\hspace{5mm} In the decoding process, the root node of $G_l$ corresponds to any variable node $v$ and $y_0 = 1$ (or $y_1 = 0$) i.e., the probability of having zero in each leaf node is one, because in the beginning of the decoding, no message symbol is decoded yet. In order to model the decoding process
via this And-Or tree we need to compute the distributions $\alpha(x)$ and $\beta(x)$ to stochastically characterize the number of children of OR and AND nodes. Luckily, this computation turns out to be easy and it corresponds to the edge perspective degree distributions of standard LDPC codes \cite{Amin2}.

\hspace{5mm} We already know the check symbol degree distribution $\Omega(x)$. The following argument establishes the coefficients of the variable node degree distribution $\lambda(x) = \sum_{d=0}^{n-1} \lambda_{d+1} x^{d+1}$.  We note that the average number of edges in $G$ is $\Omega^{\prime}(1)n$ and for simplicity we assume that the edge selections are made with replacement (remember that this assumption is valid if $k \rightarrow \infty$). Since coded symbols choose their edges uniform randomly from $k$ input symbols, any variable node $v$ having degree $d$ is given by the binomial distribution expressed as,
\begin{eqnarray}
\lambda_d = \binom{\Omega^{\prime}(1)n}{d} (1/k)^d (1 - 1/k)^{\Omega^{\prime}(1)n-d}
\end{eqnarray}
which asymptotically approaches to Poisson distribution if $\Omega^{\prime}(1)n/k$ is constant, i.e.,
\begin{eqnarray}
\lambda_d = \frac{e^{-\Omega^{\prime}(1)(1+\epsilon)} (\Omega^{\prime}(1)(1+\epsilon))^d }{d!}
\end{eqnarray}

\hspace{5mm} Now for convenience, we perform a change of variables and rewrite $\Omega(x) = \sum_{d=0}^{k-1} \Omega_{d+1} x^{d+1}$, then the edge perspective distributions are given by (borrowing ideas from LDPC code discussions)
\begin{eqnarray}
\alpha(x) = \frac{\lambda^{\prime}(x)}{\lambda^{\prime}(1)} \ \ \ \ \textrm{ and }\ \ \ \ \beta(x) = \frac{\Omega^{\prime}(x)}{\Omega^{\prime}(1)}
\end{eqnarray}

\hspace{5mm} More explicitly, we have
\begin{eqnarray}
\alpha(x) =\frac{\lambda^{\prime}(x)}{\lambda^{\prime}(1)}  &=&  \sum_{d=0}^n (d+1) \frac{e^{-\Omega^{\prime}(1)(1+\epsilon)} (\Omega^{\prime}(1)(1+\epsilon))^{d+1} x^d }{\Omega^{\prime}(1)(1+\epsilon) (d+1)!} \\
&=& e^{-\Omega^{\prime}(1)(1+\epsilon)} \sum_{d=0}^n \frac{ (\Omega^{\prime}(1)(1+\epsilon)x)^{d} }{ d!} \\
&\rightarrow& e^{-\Omega^{\prime}(1)(1+\epsilon)} e^{\Omega^{\prime}(1)(1+\epsilon)x} =  e^{\Omega^{\prime}(1)(1+\epsilon)(x-1)}  \label{Eqn1}
\end{eqnarray}
and
\begin{eqnarray}
\beta(x) &=& \frac{1}{\Omega^{\prime}(1)}\sum_{d=0}^{k-1} (d+1)\Omega_{d+1} x^d \label{Eqn2}
\end{eqnarray}

\hspace{5mm} Note that in a  standard  LT code, $\Omega(x)$ determines the message node degree distribution $\lambda(x)$. Thus, knowing $\Omega(x)$ is equivalent to knowing the asymptotic performance of the LT code using the result of lemma 3.

\hspace{5mm} Here one may be curious about the value of the limit $\lim_{l \rightarrow \infty}y_l$. The result of this limit is the appropriate unique root (i.e., the root $0 \leq x^* \leq 1$) of the following equation.
\begin{eqnarray}
\alpha(1-\beta(1-x)) = x
\end{eqnarray}
where if we insert our $\alpha(x)$ and $\beta(x)$ as found in equations (\ref{Eqn1}) and (\ref{Eqn2}), we have
\begin{eqnarray}
-\Omega^{\prime}(1)(1+\epsilon)\beta(1-x) = \ln(x)
\end{eqnarray}

\hspace{5mm} Let us make the change of variables $y=1-x$, and $k\rightarrow \infty$ i.e., $\epsilon \rightarrow 0$, we have
\begin{eqnarray}
\Omega^{\prime}(1)(1+\epsilon)\beta(y) &=& -\ln(1-y) \\
(1+\epsilon)\sum_{d=0}^{\infty} (d+1)\Omega_{d+1} y^{d} &=& \sum_{d=1}^{\infty} \frac{y^d}{d} \label{lasteqn}
\end{eqnarray}

\hspace{5mm} If we think of the limiting distribution $\Omega_d = 1/(d(d-1))$ for $d>1$ for some large maximum degree (to make $\Omega^{\prime}(1)$ a constant) and zero otherwise, it is easy to see that we can satisfy the equality asymptotically with $y \rightarrow 1$. This means at an overhead of $\epsilon \rightarrow 0$, we have $x \rightarrow 0$. Therefore the zero failure probability is possible in the limit. The letter also establishes  that LT codes used with Soliton distribution is asymptotically optimal.

\begin{figure}[t!]
\centering
\includegraphics[angle=0, height=85mm, width=115mm]{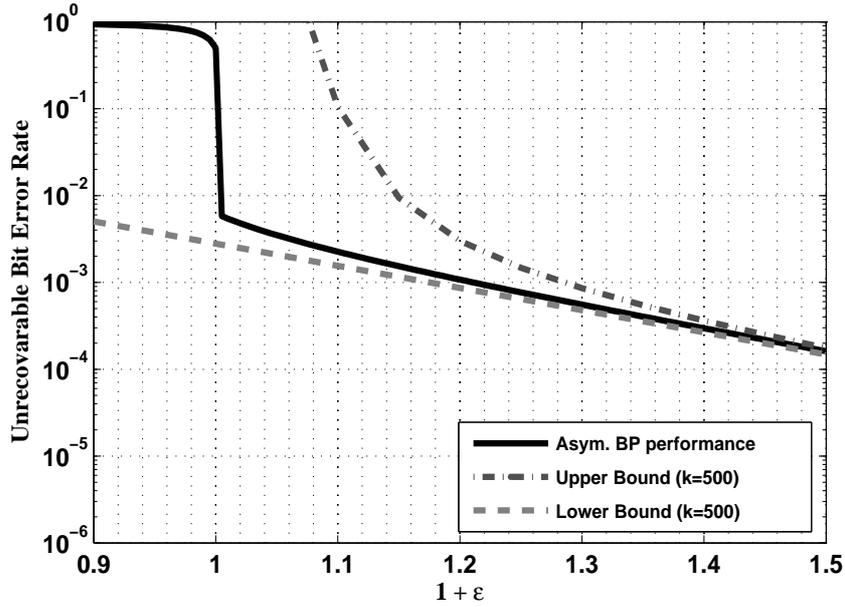}
\caption{Comparison of asymptotical BP performance and upper/lower bounds using ML decoding (assumed here $k=500$ as infinite block length is not practical for computation of these bounds) and $\widetilde{\Omega}$(x)}\label{RatelessPerF}
\end{figure}

\hspace{5mm} If $\Omega_d = 0$ for some $d > N \in \mathbb{N}$, then it is impossible to satisfy the equality above with $y=1$. This condition is usually imposed to have linear time encoding/decoding complexity for sparse fountain codes. Thus, for a given $\epsilon$ and $N$, we can solve for $y^* = 1 - x^* < 1$ ($y$ cannot be one in this case because if $y=1$, right hand side of equation (\ref{lasteqn}) will diverge whereas the left hand side will have a finite value.) and hence $1 - y^*$ shall be the limiting value of $y_l$ as $l$ tends to infinity. This corresponds to an error floor issue of the LT code using a $\Omega(x)$ as described.

\hspace{5mm} Let us provide an example code with $k=500$, an LT code with the following check symbol node degree distribution from \cite{Amin} is used with $L_{\beta} = 66$,
\begin{eqnarray}
&& \widetilde{\Omega}(x) = 0.008x + 0.494x^2 + 0.166x^3 + 0.073x^4 \nonumber \\
&&  \ \ \ \ \ \ \ \ \ \ \ \ \ \ \ \ +  0.083x^5 + 0.056x^8 + 0.037x^9 + 0.056x^{19} + 0.025x^{65} + 0.003x^{66} \label{RaptorOmega}
\end{eqnarray}
where the average degree per node is $\widetilde{\Omega}^{\prime}(1) = \approx 5.85$ which is independent of $k$. Therefore, we have a linear time encodable sparse fountain code with a set of performance curves shown in Fig. \ref{RatelessPerF}. As can be seen, asymptotical BP algorithm performance gets pretty close to the lower bound of ML decoding (It is verifiable that these bounds for ML decoding do not change too much for $k > 500$). Note the fall off starts at $\epsilon = 0$. If we had an optimal (MDS) code, we would have the same fall off but with zero error probability for all $\epsilon > 0$. This is not the case for linear time encodable LT codes.


\subsection{Expected Ripple size and Degree Distribution Design}

\hspace{5mm} We would like to start with reminding that the set of degree-one message symbols in each iteration of the BP algorithm was called the ripple. As the decoding algorithm evolves in time, the ripple size evolves as well. It can enlarge or shrink depending on the number of coded symbols released at each iteration.  The evolution of the ripple can be described by a simple random walk model \cite{Luby}, in which at each step the location jumps to next state either in positive or negative direction with probability 1/2. In a more formal definition, let $\{Z_i, i \geq 1\}$ define the sequence of independent random variables, each taking on either 1 or -1 with equal probability. Let us define $S_k = \sum_{j=1}^k Z_j$ to be the random walker on $\mathbb{Z}$. At each time step, the walker jumps one state in either positive or negative direction. It is easy to see that $\mathbb{E}(S_k) = 0$. By considering/recognazing $\mathbb{E}(Z_k^2) = 1$ and $\mathbb{E}(S_k^2) = k$, we can argue that $\mathbb{E}(|S_k|) = O(\sqrt{k})$. Moreover, using diffusion arguments we can show that \cite{Oliver}
\begin{eqnarray}
\lim_{k \rightarrow \infty} \frac{\mathbb{E}(|S_k|)}{\sqrt{k}} = \sqrt{\frac{2}{\pi}}
\end{eqnarray}
from which we deduce that the expected ripple size better be scaling with $\sim O(\sqrt{\frac{2k}{\pi}})$ as $k \rightarrow \infty$ and the ripple size resembles to the evolution of a random walk. This resemblance is quite interesting and the connections with the one dimensional diffusion process and Gaussian approximation can help us understand more about the ripple size evolution.

\hspace{5mm} From lemma 3 and subsequent analysis of the previous subsection, the standard tree analysis allows us to compute the probability of the recovery failure of an input symbol at the $l$-th iteration by
\begin{eqnarray}
y_{l+1} &=& \alpha(1-\beta(1-y_{l})) \ \ \ \textrm{for} \ \ \ l =0,2,\dots \\
&=& e^{-\Omega^{\prime}(1)(1+\epsilon)(\beta(1-y_{l}))}
\end{eqnarray}

\hspace{5mm} Assuming that the input symbols are independently recovered with probability $1-y_l$ at the $l$-th iteration (i.e., the number of recovered symbols are binomially distributed with parameters $k$ and $1-y_l$), expected number of unrecovered symbols are $ky_l$ and $ky_{l+1} = ke^{-\Omega^{\prime}(1)(1+\epsilon)(\beta(1-y_{l}))} = ke^{-\Omega^{\prime}(1-y_{l})(1+\epsilon))}$, respectively, where we used the result that $\beta(x) = \frac{\Omega^{\prime}(x)}{\Omega^{\prime}(1)}$. For large $k$,  the expected ripple size is then given by $y_lk - y_{l+1}k  = k(y_l - e^{-\Omega^{\prime}(1-y_{l})(1+\epsilon))})$. In general, we express an $x$-fraction of input symbols unrecovered at some iteration of the BP algorithm. With this new notation, the expected ripple size is then given by
 \begin{eqnarray}
k \left(x - e^{-\Omega^{\prime}(1-x)(1+\epsilon))}\right)
\end{eqnarray}

\hspace{5mm}Finally, we note that since the  actual number of unrecovered symbols converges to its expected value in the limit, this expected ripple size is the actual ripple size for large $k$\footnote{This can be seen using standard Chernoff bound arguments for binomial/Poisson distributions.}. Now, let us use our previous random walker argument for the expected unrecovered symbols $xk$ such that we make sure the expected deviation from the mean of the walker (ripple size in our context) is at least $\sqrt{2xk/\pi}$. Assuming that the decoder is required to recover $1-\gamma$ fraction of the $k$ message symbols with an overwhelming probability, then this can be formally expressed as follows,
\begin{eqnarray}
x - e^{-\Omega^{\prime}(1-x)(1+\epsilon))} \geq \sqrt{\frac{2x}{k\pi}} \label{EqnIneq0}
\end{eqnarray}
for $x \in [\gamma, 1]$. From equation  (\ref{EqnIneq0}), we can lower bound the derivative of the check node degree distribution,
\begin{eqnarray}
\Omega^{\prime}(1-x) \geq \frac{-\ln(x - \sqrt{\frac{2x}{k\pi}})}{1 + \epsilon} \label{EqnIneq}
\end{eqnarray}
for $x \in [\gamma, 1]$, or equivalently
\begin{eqnarray}
\Omega^{\prime}(x) \geq \frac{-\ln(1- x - \sqrt{\frac{2x}{k\pi}})}{1 + \epsilon} \label{EqnIneq2}
\end{eqnarray}
for $x \in [0, 1-\gamma]$ as derived in \cite{Amin}. Let us assume that the check node degree distribution is given by the limiting distribution derived earlier i.e., $\Omega(x) = (1-x)\ln(1-x) + x $ for $x \in [0,1]$ and $\gamma = 0$. It remains to check,
\begin{eqnarray}
(1-x)^{(1+\epsilon)} + x + \sqrt{\frac{2x}{k\pi}} \leq 1
\end{eqnarray}

\hspace{5mm} If we assume $k \rightarrow \infty$, we shall have
\begin{eqnarray}
(1-x)^{(1+\epsilon)} \leq 1-x
\end{eqnarray}
which is always true for $x \in [0,1]$ and for any positive number $\epsilon > 0$. Thus, the limiting distribution conforms with the assumption that the ripple size is evolving according to a simple random walker.  For a given $\gamma$,  a way to design a degree distribution satisfying equation  (\ref{EqnIneq2}) is to discretize the interval $[0, 1-\gamma]$ using some $\Delta \gamma > 0$ such that we have a union of multiple disjoint sets, expressed as
\begin{eqnarray}
[0, 1-\gamma] = \bigcup_{i=0}^{(1-\gamma)/\Delta \gamma -1} \left[i\Delta\gamma \ \ \ (i+1)\Delta\gamma\right),
\end{eqnarray}

\hspace{5mm} We require that equation  (\ref{EqnIneq2}) holds at each discretized point in the interval $[0, 1-\gamma]$, which eventually gives us a set of inequalities involving the coefficients of $\Omega(x)$. Satisfying these set of inequalities we can find possibly more than one solution for $\{\Omega_d, d=1,2,\dots,k\}$ from which we choose the one with minimum $\Omega^{\prime}(1)$. This is similar to the linear program whose procedure is outlined in \cite{Amin}, although the details of the optimization is omitted.

\hspace{5mm} We evaluate the following inequality at $M = (1-\gamma)/\Delta \gamma + 1$ different discretized points,
\begin{eqnarray}
-\Omega^{\prime}(x) = \sum_{d=1}^F c_d x^{d-1} \leq \frac{\ln\left(1- x - \sqrt{\frac{2x}{k\pi}}\right)}{1 + \epsilon} \label{Opteqn1}
\end{eqnarray}
where $F$ is the maximum degree of $\Omega(x)$ and $\Omega_d = -c_d / d$. Let us define $\textbf{c} = [c_1, c_2, \dots , c_F]^T$, $\textbf{b} = [b_1, b_2, \dots , b_M]^T$ where $b_i $ is the right hand side of equation (\ref{Opteqn1}) evaluated at $x = (i-1) \Delta \gamma$. For completeness, we also assume a lower bound vector $\textbf{lc} = [-1, -2, \dots , -F]^T$ for $\textbf{c}$. Let $\textbf{A}$ be the $M \times F$ matrix such that $A_{i,j} = (i\Delta \gamma)^{j}$ for $i=0,1,\dots,M-1$ and $j=0,1,\dots,F-1$. Thus, we have the following minimization problem\footnote{A \textsc{Matlab} implementation of this linear program can be found at $http://suaybarslan.com/optLTdd.txt$.} to solve
\begin{eqnarray}
\min_{c_1,c_2,\dots,c_F} -\textbf{1}^T\textbf{c} \textmd{  OR } \max_{c_1,c_2,\dots,c_F} \textbf{1}^T\textbf{c} \ \ \ \textrm{such that } \ \ \  \textbf{A}\textbf{c} \leq \textbf{b},  \textbf{lc} \leq \textbf{c} \leq \textbf{0}  \ \textrm{and} \ \Omega(1) = 1.
\end{eqnarray}

\hspace{5mm} Note that the condition $\Omega(1) = 1$ imposes a tighter constraint than does $\textbf{lc} \leq \textbf{c} \leq \textbf{0}$ and is needed to assure we converge to a valid probability distribution $\Omega(x)$. The degree distribution in equation (\ref{RaptorOmega}) is obtained using a similar optimization procedure for $k=65536$ and $\epsilon = 0.038$ in \cite{Amin}. Other constraints such as $\Omega^{\prime\prime}(0) = 1$ are possible based on the design choices and objectives.

\hspace{5mm} Let us choose the parameter set \{$\gamma = 0.005$, $\Delta \gamma = 0.005$\} and various $k$  and corresponding $\epsilon$ values as shown in Table \ref{TableOpt}. The results are shown for $k=4096$ and $k=8912$. As can be seen the probabilities resemble to a Soliton distribution whenever $\Omega_i$ is non-zero. Also included are the average degree numbers per coded symbol for each degree distribution.

\begin{table}
\begin{center}
    \begin{tabular}{ | l | l | l |}
    \hline
    \textbf{k} & \textbf{4096} & \textbf{8192}  \\ \hline \hline
    $\boldsymbol{\Omega_1}$ &  0.01206279868062 & 0.00859664884231  \\ \hline
    $\boldsymbol{\Omega_2}$ & 0.48618222931140 &  0.48800207839031  \\ \hline
    $\boldsymbol{\Omega_3}$  & 0.14486030215468 & 0.16243601073478  \\ \hline
    $\boldsymbol{\Omega_4}$ &  0.11968155126998 & 0.06926848659608  \\ \hline
    $\boldsymbol{\Omega_5}$ & 0.03845536920060 & 0.09460770077248  \\ \hline
    $\boldsymbol{\Omega_8}$ &  0.03045905002768 &  \\ \hline
    $\boldsymbol{\Omega_9}$ & 0.08718444024457 &  0.03973381508374  \\ \hline
    $\boldsymbol{\Omega_{10}}$  &   &   0.06397077147921  \\ \hline
    $\boldsymbol{\Omega_{32}}$ & 0.08111425911047 &   \\ \hline
    $\boldsymbol{\Omega_{34}}$ &  & 0.06652107350334  \\ \hline
    $\boldsymbol{\Omega_{35}}$ &  & 0.00686341459082  \\ \hline \hline
    $\boldsymbol{\epsilon}$ & 0.04 & 0.03 \\ \hline \hline
    $\boldsymbol{\Omega^{\prime}(1)}$ & 5.714 & 5.7213 \\
    \hline
    \end{tabular}
\end{center}
\caption{An example optimization result for $k=4096$ and $k=8192$.} \label{TableOpt}
\end{table}

\subsection{Systematic Constructions of LT codes}

\hspace{5mm} The original construction of LT codes were based on a non-systematic form i.e., the information symbols that are encoded are not part of the codeword. However, proper design of the generator matrix $\textbf{G}$ allows us to construct efficiently systematic fountain codes. This discussion is going to be given emphasis in the next section when we discuss concatenated fountain codes.

\subsection{Generalizations/Extensions of LT codes: Unequal Error Protection}

\hspace{5mm} There have been three basic approaches for the generalization of LT codes which are detailed in references \cite{Rahnavard2}, \cite{Sejdinovic} and \cite{Arslan}. In effect, LT code generalizations are introduced to provide distinct recovery properties associated with each symbol in the message sequence. All three approaches are similar in that they subdivide the message symbols into $r$ disjoint sets $s_1, s_2, \dots, s_r$ of sizes $\alpha_1s, \alpha_2s, \dots, \alpha_rs$, respectively, such that $\sum_{i=1}^r\alpha_i = 1$. A coded symbol is generated by first selecting a node degree according to a suitable degree distribution, then edges are selected (unlike original LT codes) non-uniform randomly from the symbols contained in sets $s_1, s_2, \dots, s_r$. This way, number of edge connections per set shall be different and thus different sets will have different recovery probabilities under BP decoding algorithm. In \cite{Rahnavard2}, such an idea is used to give unequal error protection (UEP) as well as unequal recovery time (URT) properties to the associated LT code. With URT property, messages are recovered with different probabilities at different iterations of the BP algorithm. In other words, some sets are recovered early in the decoding algorithm than are the rest of the message symbols. Later, a similar idea is used to provide UEP and URT based on expanding windowing techniques \cite{Sejdinovic}. More specifically, window $j$ is defined  as $W_j \triangleq \bigcup_{i=1}^j s_i$ and window selections are performed first before the degrees for coded symbols are selected. Therefore, $r$ different degree distributions are used instead of a unique degree distribution. After selecting the appropriate expanded window, edge selections are performed uniform randomly. Expanding window fountain (EWF) codes are shown to be more flexible and therefore they demonstrate better UEP and URT properties.

\hspace{5mm} Lastly, a generalization of both of these studies is conducted in \cite{Arslan}, in which the authors performed the set/window selections based on the degree number of the particular coded symbol. That is to say, this approach selects the degrees according to some optimized degree distribution and then make the edge selections based on the degree number. This way, a more flexible coding scheme is obtained. This particular generalization leads however to many more parameters subject to optimization, particularly with a set of application dependent optimization criteria.

\hspace{5mm} Let us partition the information block
into $r$ variable size disjoint sets $s_1,s_2,\dots,s_r$ ($s_j$ has
size $\alpha_jk, j=1,\dots,r$ such that $\sum_{j=1}^r \alpha_j =1$
and the $\alpha_jk$ values are integers). In the encoding process,
after choosing the degree number for each coded symbol, authors select
the edge connections according to a distribution given by

\hspace{5mm} \textbf{Definition 2:} \emph{Generalized Weighted Selection Distribution}.
\begin{enumerate}
\item[$\bullet$] For
$i=1,\dots,k$, let $P_{i}(x) = \sum_{j=1}^{r}p_{j,i} x^j$
where $p_{j,i} \geq 0$ is the conditional probability of choosing
the
 information set $s_j$,  given that the degree of the coded symbol is
 $i$ and $\sum_{j=1}^r p_{j,i}=1$.
\end{enumerate}

\hspace{5mm} Note that $p_{j,i}$ are design parameters of the system, subject to
optimization. For convenience,  authors denote the proposed selection distribution in a
matrix form as follows:
\[ \textbf{P}_{r \times k}  = \left[ \begin{array}{cccc}
p_{1,1} & p_{1,2} & \dots & p_{1,k} \\
p_{2,1} & p_{2,2} & \dots & p_{2,k} \\
\vdots & \vdots & \dots & \vdots \\
p_{r-1,1} & p_{r-1,2} & \dots & p_{r-1,k} \\
p_{r,1} & p_{r,2} & \dots & p_{r,k}
 \end{array} \right]\]

\hspace{5mm} Since the set of probabilities in each column sums to unity, the
number of design parameters of $\textbf{P}_{r \times k}$ is
$(r-1)\times k$. Similarly, the degree distribution  can be
expressed in a vector form as $\boldsymbol{\Omega}_k$, where the
$i$th vector entry is the probability that a coded symbol chooses
degree $i$ i.e., $\Omega_i$. Note $\boldsymbol{\Omega}_k$ and $\textbf{P}_{r
\times k}$ completely determine the performance of the proposed generalization.

\hspace{5mm} In the BP algorithm, we observe that not all the check
nodes decode information symbols at each iteration. For example,
degree-one check nodes immediately decode neighboring information
symbols at the very first iteration. Then, degree two
and three check nodes  recover some of the information bits later in the sequence of iterations. In general, at the
later update steps of iterations, low degree check nodes will
already be released from the decoding process,  and higher degree
check nodes start decoding the information symbols (due to edge
eliminations). So the coded symbols take part in different stages of
the BP decoding process depending on their degree numbers.

\hspace{5mm} UEP and URT is achieved
by allowing coded symbols to make more edge connections with more
important information sets. This increases the probability of
decoding the more important symbols. However,  coded symbols are
able to decode information symbols in different iterations of the BP
depending on their degree numbers. For example, at the second
iteration of the BP algorithm, the probability that degree-two coded
symbols decode information symbols is higher than that of coded
symbols with degrees larger than two\footnote{This observation will be quantified in Section 5 by drawing connections to random graph theory.}. If the BP algorithm stops unexpectedly
at early iterations, it is essential that the more important
information symbols are recovered. This suggests that it is
beneficial to have low degree check nodes generally make edge
connections with important information sets. That is the idea behind this generalization.

\hspace{5mm} In the encoding process of the generalization for EWF codes, after choosing the
degree number for each coded symbol, authors select the edge connections
according to a distribution given by

\hspace{5mm} \textbf{Definition 3:} \emph{Generalized Window Selection Distribution}.
\begin{enumerate}
\item[$\bullet$] For
$i=1,\dots,k$, let $L_{i}(x) = \sum_{j=1}^{r}\gamma_{j,i} x^j$
where $\gamma_{j,i} \geq 0$ is the conditional probability of
choosing the
 $j$-th window $W_j$,  given that the degree of the coded symbol is
 $i$ and $\sum_{j=1}^r \gamma_{j,i}=1$.
 \end{enumerate}

\hspace{5mm} Similar to the previous generalization, $ \gamma_{j,i}$ are design
parameters of the system, subject to optimization. For convenience,
we  denote the proposed window selection distribution in a matrix form as follows:
\[ \textbf{L}_{r \times k}  = \left[ \begin{array}{cccc}
\gamma_{1,1} & \gamma_{1,2} & \dots & \gamma_{1,k} \\
\gamma_{2,1} & \gamma_{2,2} & \dots & \gamma_{2,k} \\
\vdots & \vdots & \dots & \vdots \\
\gamma_{r-1,1} & \gamma_{r-1,2} & \dots & \gamma_{r-1,k} \\
\gamma_{r,1} & \gamma_{r,2} & \dots & \gamma_{r,k}
 \end{array} \right]\]

 \hspace{5mm} The set of probabilities in each column sums to unity, and the
number of design parameters of $\textbf{L}_{r \times k}$ is again
$(r-1)\times k$. Similarly, we observe that
$\boldsymbol{\Omega}_{k} $ and $\textbf{L}_{r \times k}$ completely
determine the performance of the proposed generalization of EWF
codes. Authors proposed a method for reducing the set of parameters of these generalizations for a progressive source transmission scenario. We refer the interested reader to the original studies \cite{Arslan} and \cite{Arslan2}.


\section{\textbf{Concatenated Linear Fountain Codes}}

\hspace{5mm} It is wroth mentioning that the name ``concatenated fountain code" might not be the best choice, however many realizations of this class of codes go with their own name in literature, making the ideal comparison platform hardly exist. Therefore, we shall consider them under the name of ``concatenated fountain codes" hereafter.

\begin{figure}[t!]
\centering
\includegraphics[angle=0, height=45mm, width=105mm]{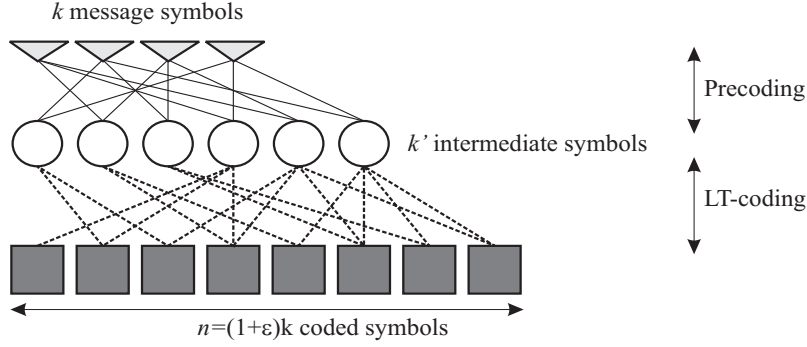}
\caption{Concatenated fountain codes consist of one or more precoding stages before an LT type encoding at the final stage of the encoder is used.}\label{probcomp}
\end{figure}

\hspace{5mm} In our previous discussion, we have seen that it is \underline{not possible} to have the whole source block decoded with negligible error probability, using an LT code in which the number of edges of the decoding graph are scaling linearly with the number of source symbols. The idea of concatenated fountain codes is to concatenate the linear-time encodable LT code that can recover $1-\gamma$ fraction of the message symbols (with high probability) with one or more stages of linear-time encodable fixed rate erasure correcting code/s given that the latter operation (coding) establishes the full recovery of the message with high probability.  Therefore, the overall linear-time operation is achieved by preprocessing (precoding) the message symbols in an appropriate manner. This serial concatenation idea is  depicted in Fig. \ref{probcomp}. As can be seen, $k$ source symbols are first encoded into $k^\prime$ intermediate symbols using a precode. The the precode codeword is encoded using a special LT code to generate $n = (1+\epsilon)k$ coded symbols.

\hspace{5mm} Here the main question is to design the check node degree distribution (for the LT code) that will allow us to recover at least $1-\gamma$ fraction of the message symbols with overwhelming  probability. We already have given an argument and a linear program methodology in subsection 2.4  in order to find a suitable class of distributions. However, those arguments assumed finite values for $k$ to establish the rules for the design of an appropriate degree distribution. In addition, optimization tools might be subject to numerical instabilities which must be taken care of. Given these challenges, in the following, we  consider asymptotically good check node degree distributions. Note that such asymptotically suitable degree distributions can still be very useful in practice whenever the message block size $k$ is adequately large.

\subsection{Asymptotically good degree distributions}

\hspace{5mm} Resorting to the previous material presented in Section 3.4, a check symbol degree distribution $\Omega(x)$ must conform with the following inequality for a given $\gamma > 0$ in order to be part of the class of linear-time encodable concatenated fountain codes.
\begin{eqnarray}
e^{-\Omega^{\prime}(1-x)(1+\epsilon)} &<& x \ \ \textrm{for} \ \ x\in[\gamma,1] \label{KeyEqn}
\end{eqnarray}

\hspace{5mm} In the past (pioneering works include \cite{Petar} and \cite{Amin}),  asymptotically good (but not necessarily optimal) check node degree distributions that allow linear-time LT encoding/decoding are proposed based on the Soliton distribution. In otherwords, such good distributions are developed based on an instance of a Soliton distribution with a maximum degree $F$. More specifically let us assume $\Pi(x) = \sum_{d=1}^F \eta_d x^d = 1/F + \sum_{d=2}^F \frac{1}{d(d-1)} x^d$ be the modified Soliton distribution assuming that the source block is of size $> F$ (from equation (\ref{SD1})). Both in Online codes \cite{Petar} and Raptor codes \cite{Amin}, a generic degree distribution of the following form is assumed,
\begin{eqnarray}
\Omega_F(x) = \sum_{d=1}^F c_d \eta_d x^d \label{GeneralEqn}
\end{eqnarray}

 \hspace{5mm}Researchers designed $F$ and the coefficient set $\{c_i\}_{i=1}^{F}$ such that from any $(1 + \epsilon)k$ coded symbols, all the message symbols but a $\gamma$ fraction can correctly be recovered with overwhelming probability. More formally, we have the following theorem that establishes the conditions for the existence of asymptotically good degree distributions.

\hspace{5mm} \textbf{Theorem 4:} \emph{ For any message of size $k$ blocks and for parameters $\epsilon > 0$ and a BP decoder error probability $\gamma > 0$, there exists a distribution $\Omega_F(x)$ that can recover a $1-\gamma$ fraction of the original message from any $(1+\epsilon)k$ coded symbols in time proportional to $k \ln(g(\epsilon,\gamma))$ where the function $g(\epsilon,\gamma)$ depends on the choice of $F$.
}

\hspace{5mm} The proof of this theorem depends on the choice of $F$ as well as the coefficients $\{c_i\}_{i=1}^{F}$. For example, it is easy to verify that for online codes these weighting coefficients are given by $c_1 = (\epsilon F -1)/(1+\epsilon)$ and $c_i = \frac{F+1}{(1+\epsilon)(F-1)}$ for $i=2,3,\dots,F$. Similarly, for Raptor codes it can be shown that $c_1 = \frac{\mu F}{\mu + 1}$, $c_i = \frac{1}{\mu + 1}$ for $i=2,3,\dots,F-1$ and $c_F = \frac{F}{\mu + 1}$ for some $\mu \geq \epsilon$. In that respect, one can see the clear similarity of two different choices of the asymptotically good degree distribution. This also shows the non-uniqueness of the solution established by the two independent studies in the past.

\hspace{5mm} PROOF of Theorem 4: Since both codes show the existence of one possible distribution that is a special case of the general form given in equation (\ref{GeneralEqn}), we will give one such appropriate choice and then prove this theorem by giving the explicit form of $g(\epsilon,\gamma)$. Let us rewrite the condition of (\ref{KeyEqn}) using the degree distribution $\Omega_F(x)$,
\begin{align}
-\Omega^{\prime}_F(x)(1+\epsilon) &< \ln(1-x) \ \ \textrm{for} \ \ x\in(0, 1-\gamma]  \\
\left(\frac{c_1}{F} + \sum_{i=1}^{F-1} \frac{c_{i+1}}{i}x^{i} \right)(1+\epsilon) &>  -\ln(1-x) \ \ \textrm{for} \ \ x\in(0, 1-\gamma]
\end{align}

\hspace{5mm}  Let us consider the coefficient set i.e., $c_1$ and $c_i = c$, i.e., some constant $c$ for $i=2,3,\dots,F$. In fact by setting $\Omega_F(1) = 1$, we can express $c = \frac{F - c_1}{F - 1}$. We note again that such a choice might not be optimal but sufficient to prove the asymptotical result.
\begin{align}
\Omega^{\prime}_F(x) &= \frac{c_1}{F} + \frac{F - c_1}{F - 1} \sum_{i=1}^{F-1} \frac{x^i}{i} \\
&= \frac{c_1}{F} + \frac{F - c_1}{F - 1} \left( \sum_{i=1}^{\infty} \frac{x^i}{i} - \sum_{i=F}^{\infty} \frac{x^i}{i} \right) \\
&= \frac{c_1}{F} - \frac{F - c_1}{F - 1} \left( \ln(1-x) + \sum_{i=F}^{\infty} \frac{x^i}{i} \right)
\end{align}

\hspace{5mm} We have the following inequality,
\begin{align}
\frac{c_1}{F} - \frac{F - c_1}{F - 1} \left( \ln(1-x) + \sum_{i=F}^{\infty} \frac{x^i}{i} \right) > \frac{-\ln(1-x)}{1+\epsilon}
\end{align}
or equivalently,
\begin{align}
\frac{c_1}{F} - \frac{F - c_1}{F - 1}  \sum_{i=F}^{\infty} \frac{x^i}{i}  > \left( \frac{F - c_1}{F - 1} - \frac{1}{1+\epsilon} \right) \ln(1-x) \label{Inequality1}
\end{align}

\hspace{5mm} Assuming that the righthand side of the inequality (\ref{Inequality1}) is negative, with $x\in(0, 1-\gamma]$ we can upper bound $c_1$ as follows
\begin{align}
 \frac{\epsilon F +1}{\epsilon + 1} >   c_1 \label{Ubc1}
\end{align}
and similarly, assuming that the lefthand side of the inequality (\ref{Inequality1}) is positive, we can lower bound $c_1$ as follows,
\begin{align}
c_1 > \frac{ F^2  \sum_{i=F}^{\infty} x^i/i }{ F-1 + F  \sum_{i=F}^{\infty} x^i/i } \label{lbc1}
\end{align}

\hspace{5mm} For example, for online codes, $c_1 = (\epsilon F -1)/(1+\epsilon)$ and for Raptor codes, $c_1 = \epsilon F/(1+\epsilon)$ are selected where both choices satisfy the inequality (\ref{Ubc1}). If we set $c_1 = (\epsilon F -1)/(1+\epsilon)$, using the inequality (\ref{lbc1}) we will have for large $F$,
\begin{align}
\frac{(\epsilon F - 1)(F-1)}{F(F+1)} \approx \frac{\epsilon F - 1}{F} > \sum_{i=F}^{\infty} \frac{x^i}{i} \label{IneqF1}
\end{align}

\hspace{5mm} Also if we set  $c_1 = \epsilon F/(1+\epsilon)$, we similarly will reach at the following inequality
\begin{align}
\frac{\epsilon(F-1)}{F}  > \sum_{i=F}^{\infty} \frac{x^i}{i} \label{IneqF2}
\end{align}

\hspace{5mm} From the inequalities (\ref{IneqF1}) and/or (\ref{IneqF2}), we can obtain lower bounds on $F$ provided that $\sum_{i=F}^{\infty} x^i/i < \xi$,
\begin{align}
F > \frac{1}{\epsilon - \xi} \ \ \ \textrm{or} \ \ \ F > \frac{\epsilon}{\epsilon - \xi}, \label{Condition1}
\end{align}
respectively. For online codes, the author set $F = \frac{\ln(\gamma) + \ln(\epsilon/2)}{\ln(1-\gamma)}$ whereas the author of Raptor codes set
\begin{align}
F = \left\lfloor 1/\gamma + 1 \right\rfloor + 1  \ \ \ \textrm{with} \ \ \ \gamma = \frac{\epsilon/2}{1+2\epsilon}
\end{align}

\hspace{5mm} Therefore given $\epsilon, \gamma$ and $x\in(0, 1-\gamma]$, as long as the choices for $F$ comply with the inequalities (\ref{Condition1}), we obtain an asymptotically good degree distribution that proves the result of the theorem. We note here that the explicit form of $g(\epsilon,\gamma)$ is given by $F$ and its functional form in terms of $\epsilon$ and $\gamma$.

\hspace{5mm} Furthermore, using both selections of $F$ will result in the number of edges given by
\begin{eqnarray}
k \sum_{d=1}^F d c_d \pi_d  &\approx&  \frac{k\epsilon}{1 + \epsilon} +  \frac{k}{1+\epsilon} \sum_{d=2}^{F-1} \frac{1}{d-1} + \frac{kc_F}{F-1} \\
&\approx& k \ln (F) = k \ln (g(\epsilon,\gamma)) \\
&=& O(k\ln(1/\epsilon))
\end{eqnarray}
which is mainly due to the fact that for a given $\epsilon$, $F$ can be chosen in proportional to $1/\epsilon$ to satisfy the asymptotical condition. This result eventually shows that choosing an appropriate $F$, the encoding/decoding complexity can be made linear in $k$. \hfill  $\blacksquare$

\subsection{Precoding}

\hspace{5mm} Our main goal when designing good degree distributions for concatenated fountain codes was to dictate a $\Omega_F(x)$ on the BP algorithm to ensure the recovery of $(1-\gamma)$ faction of the intermediate message symbols with overwhelming probability. This shall require us to have an $(1-\gamma)$ rate $(k^{\prime},k)$ MDS type code which can recover any $k^{\prime}-k$ or less erasures with probability one where $\gamma = 1 - k/k^{\prime}$. This is in fact the ideal case (capacity achieving) yet lacks the ease of implementation. For example, we recall that the best algorithm that can decode MDS codes based on algebraic constructions (particularly Reed-Solomon(RS) codes) takes around $O(k \log k \log \log k)$ operations \cite{Ron}. This time complexity is usually not acceptable when the data rates are of the order of Mbps \cite{Byers}. In addition, manageable complexity RS codes are of small block lengths such as $255$ (symbols) which requires us to chop bigger size files into fixed length chunks before operation. This may lead to interleaving overhead \cite{QualcommRaptor}.

\hspace{5mm}  In order to maintain the overall linear time operation, we need linear time encodable precodes. One design question is whether we can use multiple precoding stages. For example, tornado codes would be a perfect match if we allow a cascade of precoding stages while ensuring the recovery of the message symbols from $1-\gamma$ fraction of the intermediate symbols as the block length tends large.

\begin{figure}[t!]
\centering
\includegraphics[angle=0, height=60mm, width=145mm]{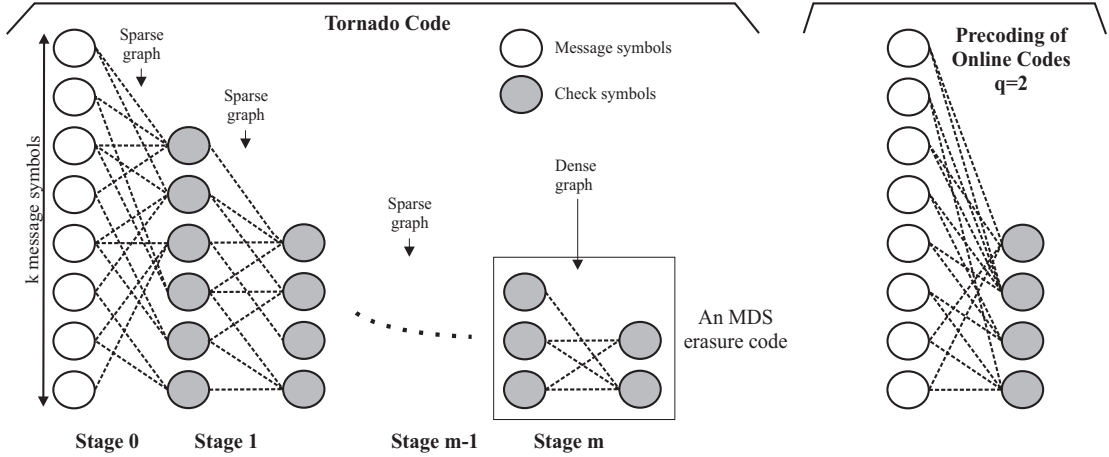}
\caption{A tornado code consists of a cascade of sparse graphs followed by a dense graph with the MDS property.}\label{tornado}
\end{figure}

\subsubsection{Tornado Codes as Precodes}

\hspace{5mm} Tornado codes \cite{Luby3}  are systematic codes, consisting of $m+1$ stages of parity symbol generation process, where in each stage $\beta \in (0,1)$ times of the previous stage symbols are generated as check symbols. The encoding graph is roughly shown in Fig. \ref{tornado}. In stage $0$, $\beta k$ check symbols are produced from $k$ message symbols. Similarly in stage 1, $\beta^2 k$ check symbols are generated and so on so forth. This sequence of check symbol generation is truncated by an MDS erasure code of rate $1-\beta$. This way, the total number of check symbols so produced are given by
\begin{eqnarray}
\sum_{i=1}^{m} \beta^i k + \frac{\beta^{m+1}k}{1 - \beta} = \frac{\beta k }{1 - \beta}
\end{eqnarray}

\hspace{5mm}Therefore with $k$ message symbols, the block length of the code is $k + \beta k / (1 - \beta)$ and the rate is $1 - \beta$. Thus, a tornado code encodes $k$ message symbols into $(1 + \beta + \beta^2 + \dots) k$ coded symbols. To maintain the linear time operation, authors chose $\beta^m k = O(\sqrt{k})$ so that the last MDS erasure code has encoding and decoding complexity linear in $k$ (one such alternative for the last stage could be a Cauchy Reed-Solomon Code \cite{Bloemer}).

\hspace{5mm}The decoding operation starts with decoding the last stage ($m$-th stage) MDS code. This decoding will be successful if at most $\beta$ fraction of the last $\beta^m k / (1-\beta)$ check symbols have been lost. If the $m$-th stage decoding is successful, $\beta^m k$ check symbols are used to recover the lost symbols of the $(m-1)$th stage check symbols. If there exists a right node whose all left neighbors except single one are known, then using simple XOR-based logic, unknown value is recovered. The lost symbols of the other stages are recovered using the check symbols of the proceeding stage in such a recursive manner.
For long block lengths, it can be shown that this $1-\beta$ rate tornado code can recover an average $\beta (1-\epsilon)$ fraction of lost symbols using this decoding algorithm with high probability in time proportional to $k \ln(1/\epsilon)$. The major advantage of the cascade is to enable linear time operation on the encoding and decoding algorithms although the practical applications use few cascades and thus the last stage input symbols size is usually greater than $O(\sqrt{k})$. This choice is due to the fact that asymptotical results assume erasures to be distributed over the codeword uniform randomly. In fact in some of the practical scenarios, erasures might be quite correlated and bursty.

\hspace{5mm} \textbf{Exercise 5:} Show that the code rate of the all cascading stages but the last stage of the tornado code has the rate $> 1 - \beta$.

\hspace{5mm} The following precoding strategy is proposed within the online coding context \cite{Petar}. It exhibits similarity to tornado codes and will therefore be considered as a special case of tornado codes (See Fig. \ref{tornado}). The main motivation for this precoding strategy is to increase the number of edges slightly in each of the cascade  such that a single stage tornado code can practically be sufficient to recover the residual lost intermediate symbols. More specifically, this precoding strategy encodes $k$ message symbols into $(1 + q\beta) k$ coded symbols such that the original message fails to be recovered completely with a constant probability proportional to $\beta^q$. Here, each message symbol has a fixed degree $q$ and chooses its neighbors uniform randomly from the $q\beta k$ check symbols. The decoding is exactly the same as the one used for tornado codes. Finally, It is shown in \cite{Petar} that a missing random $\beta$ fraction of the original $k$ message symbols can be recovered from a random $1-\beta$ fraction of the check symbols with success probability $1- \beta^{q}$.

\begin{figure}[t!]
\centering
\includegraphics[angle=0, height=57mm, width=62mm]{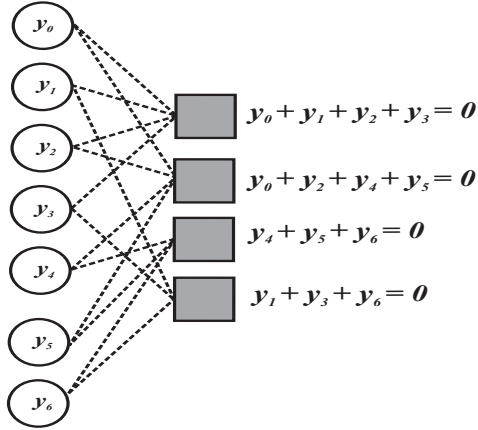}
\caption{An irregular (7,3) LDPC block code.}\label{ldpc}
\end{figure}

\subsubsection{LDPC Codes as Precodes}

\hspace{5mm} LDPC codes are one of the most powerful coding techniques of our age equipped with easy encoding and decoding algorithms. Their capacity approaching performance and parallel implementation potential make them one of the prominent options for precoding basic LT codes. This approach is primarily realized with the introduction of Raptor codes in which the author proposed a special class of (irregular) LDPC precodes to be used with linear time encodable/decodable LT codes. A bipartite graph also constitutes the basis for LDPC codes. Unlike their dual code i.e., fountain codes, LDPC code check nodes  constrain the sum of the neighbor values (variable nodes) to be zero. An example is shown in Fig. \ref{ldpc}. The decoding algorithm is very similar to BP decoding given for LT codes although here degree-1 variable node is not necessary for the decoding operation to commence. For BP decoder to continue at each iteration of the algorithm, there must be at least one check node with at most one edge connected to an erasure. The details of LDPC codes is beyond the scope of this note, I highly encourage the interested reader to look into the reference \cite{Amin2} for details.

\hspace{5mm} Asymptotic analysis of LDPC codes reveals that these codes under BP decoding has a threshold erasure probability $\epsilon_0^*$ \cite{Di} below which error-free decoding is possible. Therefore, an unrecovered fraction of $\gamma < \epsilon_0^*$ shall yield an error-free decoding in the context of asymptotically good concatenated fountain codes. However, practical codes are finite length and therefore a finite length analysis of LDPC codes is of great significance for predicting the performance of finite length concatenated fountain codes which use LDPC codes in their precoding stage. In the finite length analysis of LDPC codes under BP decoding and BEC, the following definition is the key.

\hspace{5mm} \textbf{Definition 4:} (\emph{Stopping Sets}) In a given bipartite graph of an LDPC code, a stopping set $S$ is a subset of variable nodes (or an element of the powerset\footnote{The power set of any set $S$ is the set of all subsets of $S$, denoted by $\mathcal{P}(S)$, including the empty set as well as $S$ itself.} of the set of variable nodes) such that every check node has zero, two or more connections (through the graph induced by $S$) with the variable nodes in $S$.

\hspace{5mm}Since the union of stopping sets is another stoping set, it can be shown that (\emph{Lemma 1.1} of \cite{Di}) any subset of the set of variable nodes has a unique maximal stopping set (which may be an empty set or a union of small stoping sets). Exact formulations exist for example for $(l,r)$-regular LDPC code ensembles \cite{Di}. Based on such, block as well as bit level erasure rates under BP decoding are derived. This is quantified in the following theorem.

\hspace{5mm} \textbf{Theorem 5:} \emph{ For a given $\gamma$ fraction of erasures, $(l,r)$-regular LDPC code with block length $n$  has the following block erasure and bit erasure probabilities after BP decoder is run on the received sequence.
\begin{eqnarray}
P_{block,e} =  \sum_{i=0}^{n} \chi_i \binom{n}{i} \gamma^i (1-\gamma)^{n-i}
\end{eqnarray}
where $\chi_i = \left( 1 - \frac{N(i,n/r,0)}{T(i,n/r,0)}\right) $ for $i \leq n/r-1$ and $1$ otherwise, whereas
\begin{eqnarray}
P_{bit,e} = \sum_{i=0}^{n} \binom{n}{i} \gamma^i (1-\gamma)^{n-i} \sum_{s=0}^{i} \binom{i}{s}  \frac{s O(i,s,n/r,0)}{n T(i,n/r,0)} \label{LDPCPerf}
\end{eqnarray}
where also
\begin{eqnarray}
T(i,n/r,0) = \binom{n}{il} (il)! , \ \ \ \ \  N(i,n/r,0) = T(i,n/r,0) - \sum_{s>0}\binom{i}{s} O(i,s,n/r,0) \\
O(i,s,n/r,0) = \sum_{j} \binom{n/r}{j} \textmd{coef} \left(((1+x)^r - 1 - rx)^j, x^{sl}\right) (sl)! N(i-s, n/r - j, jr - sl)
\end{eqnarray}
}

\hspace{5mm} The proof of this theorem can be found in \cite{Di}. ML decoding performance of general LDPC code ensembles are also considered in the same study. Later, more improvements have been made to this formulation for accuracy. In fact, this exact formulation is shown to be a little underestimator in \cite{Johnson}. The generalization of this method for irregular ensembles is observed to be hard. Yet, very good upper bounds have been found on the probability that the induced bipartite graph has a maximal stoping set of size $s$ such as in \cite{Amin}. Irregular LDPC codes are usually the defacto choice for precoding for their capacity achieving/fast fall off performance curves although they can easily show error floors and might be more complex to implement compared to regular LDPC codes.

\subsubsection{Hamming Codes as Precodes}
\hspace{5mm} Let us begin with giving some background information about Hamming codes before discussing their potential use within the context of concatenated fountain codes.

\paragraph{Background} One of the earliest, well-known linear codes is the Hamming code. Hamming codes are defined by a parity check matrix and are able to correct single bit error. For any integer $r \geq 2$, a conventional Hamming code assume a $r \times 2^r-1$ parity check matrix where columns are binary representation of numbers $1,\dots,2^r-1$ i.e., all distinct nonzero $r-$tuples. Therefore, a binary Hamming code is a $(2^r-1,2^r-1-r,3)$ code for any integer $r \geq 2$ defined over $\mathbb{F}_2^{2^r-1}$, where $3$ denotes the minimum distance of the code. Consider the following Hamming code example for $r=3$ whose columns are binary representation of numbers $1,2,\dots,7,$
\[ \textbf{H}_{ham} =  \left( \begin{array}{cccccccccc}
0 & 0 & 0 & 1 & 1 & 1 & 1 \\
0 & 1 & 1 & 0 & 0 & 1 & 1 \\
1 & 0 & 1 & 0 & 1 & 0 & 1 \end{array} \right)_{3 \times 7} \]

\hspace{5mm}Since the column permutations does not change the code's properties, we can have the following parity check matrix and the corresponding generator matrix,
\[ \textbf{H}_{ham} =  \left( \begin{array}{cccccccccc}
1 & 1 & 0 & 1 & 1 & 0 & 0 \\
1 & 1 & 1 & 0 & 0 & 1 & 0 \\
1 & 0 & 1 & 1 & 0 & 0 & 1 \end{array} \right)_{3 \times 7}
\Longleftrightarrow  \textbf{G}_{ham} =  \left( \begin{array}{cccccccccc}
1 & 0 & 0 & 0 & 1 & 1 & 1 \\
0 & 1 & 0 & 0 & 1 & 1 & 0 \\
0 & 0 & 1 & 0 & 0 & 1 & 1 \\
0 & 0 & 0 & 1 & 1 & 0 & 1 \end{array} \right)_{4 \times 7} \]
from which we can see the encoding operation is linear time. Hamming codes can be extended to include one more parity check bit at the end of each valid codeword such that the parity check matrix will have the following form,
\[ \textbf{H}_{ex} =  \left( \begin{array}{cccccccccc}
 &  &  &  &  &  &  & 0\\
 &  &  &  \textbf{H}_{ham} &  &  &  & \vdots\\
 &  &  &  &  &  &  & 0\\
1 & 1 & 1 & \dots & 1 & 1 & 1 & 1 \end{array} \right)_{r+1 \times 2^r} \]
which increases the minimum distance of the code to $4$.

\paragraph{Erasure decoding performance} In this subsection, we are more interested in the erasure decoding performance of Hamming codes. As mentioned before, $r$ is a design parameter (= number of parity symbols) and the maximum number of erasures (any pattern of erasures) that a Hamming code can correct cannot be larger than $r$. This is because if there were more than $r$ erasures, any valid two codewords shall be indistinguishable. But it is true that a Hamming code can correct any erasure pattern with 2 or less erasures. Likewise, extended Hamming code can correct any erasure pattern with 3 or less erasures. The following theorem from \cite{Zyablov} is useful for determining the number of erasure patterns of weight $\tau \leq r$ erasures that a Hamming code of length $2^r-1$ can tolerate.

\hspace{5mm} \textbf{Theorem 6:} \emph{ Let \textbf{B} be a $r \times \tau$ matrix whose columns are chosen from the columns of a parity check matrix $\textbf{H}_{ham}$ of length $2^r-1$, where $1 \leq \tau \leq r$. Then the number of matrices \textbf{B} such that} rank$(\textbf{B} ) = \tau$, \emph{is equal to
\begin{eqnarray}
B(\tau,r) = \frac{1}{\tau!} \prod_{i=0}^{\tau-1} (2^r - 2^i) \label{Thm4}
\end{eqnarray}
and furthermore the generator function for the number of correctable erasure patterns for this Hamming code is given by
\begin{eqnarray}
g(s,r) = \binom{2^r-1}{1} s + \binom{2^r-1}{2} s^2 + \sum_{\tau=3}^r \frac{s^\tau}{\tau!} \prod_{i=0}^{\tau-1} (2^r - 2^i) \label{HammingPerf}
\end{eqnarray}
}

\hspace{5mm} PROOF: The columns of  $\textbf{H}_{ham}$ are the elements of the $r-$ dimensional binary space except the all-zero tuple. The number of matrices  \textbf{B} constructed from distinct $\tau$ columns of  $\textbf{H}_{ham}$, having rank(\textbf{B}) = $\tau$, is equal to the number of different bases of $\tau-$ dimensional subspaces. If we let $\{\textbf{b}_1, \textbf{b}_2, \dots, \textbf{b}_{\tau}\}$ denote the set of basis vectors. The number of such sets can be determined by the following procedure,
\begin{itemize}
\item Select $\textbf{b}_1$ to be equal to one of the $2^r-1$ columns of $\textbf{H}_{ham}$.
\item For $i=2,3,\dots,\tau$, select $\textbf{b}_i$ such that it is not equal to previously chosen $i-1$ basis vectors i.e., $\{\textbf{b}_1, \textbf{b}_2, \dots, \textbf{b}_{i-1}\}$. It is clear that there are $2^r - 2^{i-1}$ choices for $\textbf{b}_i, i=2,3,\dots,\tau$.
\end{itemize}

 \hspace{5mm} Since the ordering of basis vectors $\textbf{b}_i$s are irrelevant, we exclude the different orderings from our calculations and we finally reach at $B(\tau,r)$ given in equation (\ref{Thm4}). Finally we notice that $ g(1,r) = \binom{2^r-1}{1}$ and  $ g(2,r) = \binom{2^r-1}{2}$  meaning that all patterns with one or two erasures can be corrected by a Hamming code. \hfill  $\blacksquare$

\hspace{5mm} Hamming codes are usually used to constitute the very first stage of the precoding of concatenated fountain codes. The main reason for choosing a conventional or an extended Hamming code as our precoder is to help the consecutive graph-based code (usually a LDPC code) with small stoping sets. For example, original Raptor codes use extended Hamming codes to reduce the effect of stopping sets of very small size  due to the irregular LDPC-based precoding \cite{Amin}.

\hspace{5mm} \textbf{Exercise 6:} It might be a good exercise to derive the equivalent result of theorem 4 for extended binary Hamming as well as $q$-ary Hamming codes. Non-binary codes might be beneficial for symbol/object level erasure corrections for future generation fountain code-based storage systems. See Section 5 to see more on this.

\subsubsection{Repeat and Accumulate Codes as Precodes}

\hspace{5mm} The original purpose of concatenating a standard LT code with accumulate codes \cite{Jin} is to make systematic Accumulate LT (ALT) codes as efficient as standard LT encoding while maintaining the same performance. It is claimed in the original study \cite{Yuan} that additional precodings (such as LDPC) applied to accumulate LT codes (the authors call it doped ALT codes) may render the overall code more ready for joint optimizations and thus result in better asymptotical performance. The major drawback however is that this precoded accumulate LT codes demonstrate only near-capacity achieving performance if the encoder has information about the erasure rate of the channel. This means that the rateless property might have not been utilized in that context. Recent developments\footnote{Please see Section 5. There have been many advancements in Raptor Coding Technology before and after the company Digital Fountain (DF) is acquired by Qualcomm in 2009. For complete history, search Qualcomm's web site and search for RaptorQ technology.} in Raptor coding technology and decoder design make this approach somewhat incompetent for practical applications.

\subsection{Concatenated Fountain Code Failure Probability}

\hspace{5mm} Let us assume that $k$ message symbols are encoded into $n_1$ intermediate symbols and $n_i$ intermediate symbols are encoded into $n_{i+1}$ intermediate symbols in the $i$-th precoding stage systematically for $i=1,2,\dots,P$. Finally, last stage LT code encodes $n_P$ precoded symbols into $(1+\epsilon)k$ coded symbols with $k < n_1 < \dots < n_P < (1+\epsilon)k$.

\hspace{5mm} Let $q_{s_i}$ denote the probability that the $i$-th precode can decode $s_i$ erasures at random. Similarly, let $p_l$ denote the probability that the BP decoder for the LT code fails after recovering exactly $l$ of $n_P$ intermediate symbols for $0 \leq l \leq n_P$. Assuming residual errors after decoding LT code and each one of the precode are randomly distributed across each codeword, we have the overall decoder failure probability given by the following theorem.

\hspace{5mm} \textbf{Theorem 7:} \emph{ Let us have a systematic concatenated fountain code as defined above with the associated parameters. The failure probability of recovering $k$ information symbols of the concatenated fountain code is given by
\begin{eqnarray}
= \sum_{l=0}^{n_P} \sum_{s_P = 0}^{n_P - l} \sum_{s_{P-1} = 0}^{n_{P-1} \left(1 - \theta_P(P-1) - \frac{l+s_P}{n_P}\right)}  \dots \sum_{s_{2}=0}^{n_2 \left( 1 - \theta_P(2) - \frac{l + s_P}{n_P} \right)} p_l \prod_{i=2}^{P}q_{s_i} \left(1 - q_{n_1 \left( 1 - \theta_P(1) - \frac{l + s_P}{n_P} \right)} \right) \nonumber
\end{eqnarray}
where $\theta_P(j) = \sum_{k = j+1}^{P-1} \frac{s_k}{n_k}$ for $2 \leq j \leq P-1$.}

\hspace{5mm} PROOF: Let us condition on the $l$ recovered symbols for $0 \leq l \leq n_P$ after BP decoding of the LT code. Also, let us condition on $s_2, \dots, s_P$ erasures that are corrected by each stage of the precoding. Note that at the $P-$th precoding stage, after decoding, there shall be $(l+s_P)$ recovered symbols out of $n_P$. Since we assume these recovered symbols are uniform randomly scattered across the codeword, the number of recovered symbols in the $(P-1)$th stage precoding (i.e., within codeword of length $n_{P-1}$) is $\frac{n_{P-1}}{n_p}(l + s_P)$. Thus, the number of additional recovered symbols at the $(P-1)$ precoding stage satisfies $0 \leq s_{P-1} \leq n_{P-1} - \frac{n_{P-1}}{n_p}(l + s_P)$. In general for $2 \leq j \leq P-1$, we have
\begin{eqnarray}
0 \leq s_j &\leq& n_j - \frac{n_j}{n_{j+1}}\left(s_{j+1} + \frac{n_{j+1}}{n_{j+2}} \left( \dots + \frac{n_{P-1}}{n_P} (l + s_P) \dots \right) \right) \\
&=& n_j \left(1 - \frac{s_{j+1}}{n_{j+1}} - \frac{s_{j+2}}{n_{j+2}} - \dots - \frac{s_{P-1}}{n_{P-1}} - \frac{l + s_{P}}{n_{P}}\right)  \\
&=& n_j \left(1 - \theta_P(j) - \frac{l + s_{P}}{n_{P}}\right) \label{CFCFE}
\end{eqnarray}
where we note that the upper bounds on $s_j$s also determine the limits of the sums in the expression.

\hspace{5mm} Since the failure probability of each coding stage is assumed to be independent, we multiply the failure probabilities all together provided that we have $l, s_P, \dots, s_2$ recovered symbols just before the last precoding stage that decodes the $k$ information symbols. In order for the whole process to fail, the last stage of the decoding cascade must fail. Conditioning on the number of erasures already corrected by the rest of the precoding and LT coding stages, the number of remaining unrecovered message symbols can be calculated using equation (\ref{CFCFE}) with $j=1$. The probability that $s_1 = n_1 \left(1 - \theta_P(1) - \frac{l + s_{P}}{n_{P}}\right)$ is given by  $q_{n_1 \left( 1 - \theta_P(1) - \frac{l + s_P}{n_P} \right)}$. Therefore, if $s_1 \not= n_1 \left(1 - \theta_P(1) - \frac{l + s_{P}}{n_{P}}\right)$ the whole process shall fail and therefore the last probability expression $1 - q_{n_1 \left( 1 - \theta_P(1)  - \frac{l + s_P}{n_P} \right)}$ follows by the product rule. Finally, we sum over $l, s_P, \dots, s_2$ to find the unconditional failure probability of the concatenated fountain code decoding process.
 \hfill  $\blacksquare$

 \hspace{5mm} As a special case $P=1$, i.e., we have only one precoding stage, the failure probability will be reduced to
 \begin{eqnarray}
 \sum_{l=0}^{n_1} p_l (1 - q_{n_1 - l})
\end{eqnarray}
where we implicitly set $s_1 = 0$. Note that this is exactly the same expression given in \cite{Amin}. Here we formulate the expression for the general case.

\hspace{5mm} Using the result of Theorem 7 and previously developed expressions like equations (\ref{HammingPerf}) and (\ref{LDPCPerf}),  the failure probability of a concatenated fountain code constituted of a Hamming, an LDPC and an LT code, can be found. If the performance of the LDPC code can be upper bounded, the result of theorem 7 can still be used to find tight upper bounds for the concatenated fountain code of interest.

\subsection{Systematic Constructions of Concantenated Fountain Codes}

\hspace{5mm} In many practical applications such as large scale data storage or error resilient continuous data delivery or streaming (see next section for details of these applications), it is preferred that the data is part of the fountain codeword i.e., generated code stream is in a systematic form. In a cloud storage application for example, if there is a single storage node or unit failure in the parity section, one can avoid encoding and decoding operations using systematic fountain codes. This can dramatically reduce the cost spent to maintain large data centers \cite{Dimakis}. So far, we covered fountain codes in their non-systematic form, it turns out that there are efficient ways to construct such codes in a systematic form. Main concern is whether this shall induce some performance loss with respect to erasure recovery of the code. Note that we constrain ourselves to low-complexity decoding alternatives such as BP algorithm in the rest of our discussion of systematic fountain codes of this subsection.

\hspace{5mm} A straightforward way to construct a systematic generator matrix is given by
\begin{eqnarray}
\textbf{G} = [\textbf{I}_{k \times k} | \textbf{P}_{k \times n-k}] \label{FormG1}
\end{eqnarray}
which however leads to significant inefficiencies regarding the overhead of the code. In order to design a fountain code with low overhead, $\textbf{P}$ matrix must have non-sparse columns for binary concatenated fountain codes. In a number of previous studies \cite{Asteris}, it is shown that this objective is achievable for non-binary fountain codes at the expense of more complex encoding and decoding operations i.e., Gaussian elimination etc. For low complexity operation, codes that operate on binary logic and an associated low complexity decoding algorithm are more desirable. Therefore the form of $G$ matrix in equation (\ref{FormG1}) is not very appropriate to achieve our design goal. The approach we adapt is to design the generator matrix similar to the way we design for non-systematic fountain coding, and turn the encoding operations upside down a bit to allow systematic encoding.

\begin{figure}[t!]
\centering
\includegraphics[angle=0, height=35mm, width=135mm]{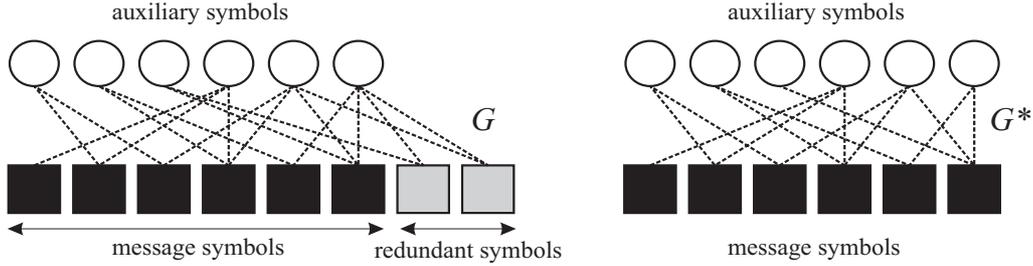}
\caption{Let $G^*$ be a BP-decodable subgraph of $G$. The fist $k$ symbols of coded symbols are assumed to be message symbols based on which we compute (using BP algorithm or back substitution) the auxiliary symbols shown as white circles based on which the redundant symbols are calculated. }\label{RatelessPer}
\end{figure}

\hspace{5mm} Let us consider a bipartite graph $G$ that generates the fountain code with $k$ source symbols and $n$ output symbols. We call a subgraph $G^* \subseteq G$ \emph{BP-decodable} if and only if the message symbols can be decoded using only the check symbols adjacent to $G^*$. Suppose that for a given check node degree distribution $\Omega(x)$, we are able to find a \emph{BP-decodable} $G^*$ with $k$ adjacent check symbols (Note that this is the minimum size of a subgraph that BP might result in a decoding success). Supposing that such $G^*$ exists,  the systematic LT encoding can be summarized as follows,
\begin{itemize}
\item Let the adjacent check symbols of $G^*$ be the user message/information symbols.
\item Run BP decoding to find the auxiliary symbol values (here due to the structure of $G^*$, we know that BP decoding is successful).
\item Generate the redundant/parity symbols as in the usual non-systematic LT encoding.
\end{itemize}

\hspace{5mm} This procedure is illustrated in Fig. \ref{RatelessPer}. Main question here is to find an appropriate $G^*$ that does not lead to any penalty in terms of overhead and show similar or better performance than the non-systematic counterpart. In \cite{Amin}, a method is suggested based on constructing a non-systematic large graph $G\supset G^{*}$ with $n$ adjacent check symbols to ensure that BP decoding is successful. Here a large graph means large overhead $\epsilon$. After finding an appropriate $G$, a search is used to identify an appropriate $G^*$ with desired properties. However, such a scheme might change the edge perspective distributions for both variable and check nodes due mainly to increased overhead, which are usually optimized for non-systematic constructions. Another viable alternative is presented in \cite{Yuan0} with smaller $\epsilon$ to keep, for example, message node degree distribution close to Poisson. A more intuitive construction of such a $G^*$ can be given. This construction assumes $\epsilon = 0$ and closely approximates the optimal edge perspective distributions. However, clear shall be from the description below that the constrained edge selection algorithm leads to dramatic changes to the variable node degree distribution and hence slightly degrade the fountain code performance.

\hspace{5mm} The algorithm consists of three basic steps as outlined below,
\\ ---------------------------------
\begin{itemize} \small
\item \textbf{Initialization:} Let the first check symbol  node degree to be one and the second node degree to be two i.e., $d_1=1$ and $d_2=2$. This is necessary for the iterative algorithm to proceed.
\item \textbf{Step 1:} Generate the degree of other check symbol nodes according to $\Omega_{*}(x) = \frac{x}{1-\Omega_1}(\frac{\Omega(x)}{x} - \Omega_1)$ where $\Omega(x)$ is the degree distribution used for the non-systematic code. Let $\textbf{V} = [1 \ 2 \ d_3 \ \dots \ d_k]$be the set of node degrees generated.
\subitem $\diamondsuit$ \textbf{Step 1 Check:} Let $d_{(i)}$ be the i-th minimum of \textbf{V} and if for all $i=1,2,\dots,k$ we have $d_{(i)} \leq i$ then we say $\textbf{V}$ can generate an appropriate $G^{*}$ and continue with Step 2. Otherwise, go back to Step 1.
\item \textbf{Step 2:} Since the degree one probability is zero with the new distribution, we can decode at most one auxiliary block per iteration.
    \subitem $\diamondsuit$ \textbf{Step 2 Edge Selections:} The edge connections are made such that in each iteration exactly one auxiliary block is recovered. This is assured by the following procedure.
    \subitem \textbf{For iteration $i=3$ to $k$}
    \subsubitem  The $i$-th minimum degree coded symbol make $d_{(i)}-1$ connections with the $i-1$ recovered auxiliary symbols and one connection with the rest of the $k-i+1$ auxiliary symbols. Note that this is possible due to \emph{Step 1 Check} and at the $i$-th iteration, there are exactly $i-1$ recovered auxiliary symbols.
\end{itemize}
\normalsize
---------------------------------
\begin{figure}[t!]
\centering
\includegraphics[angle=0, height=85mm, width=130mm]{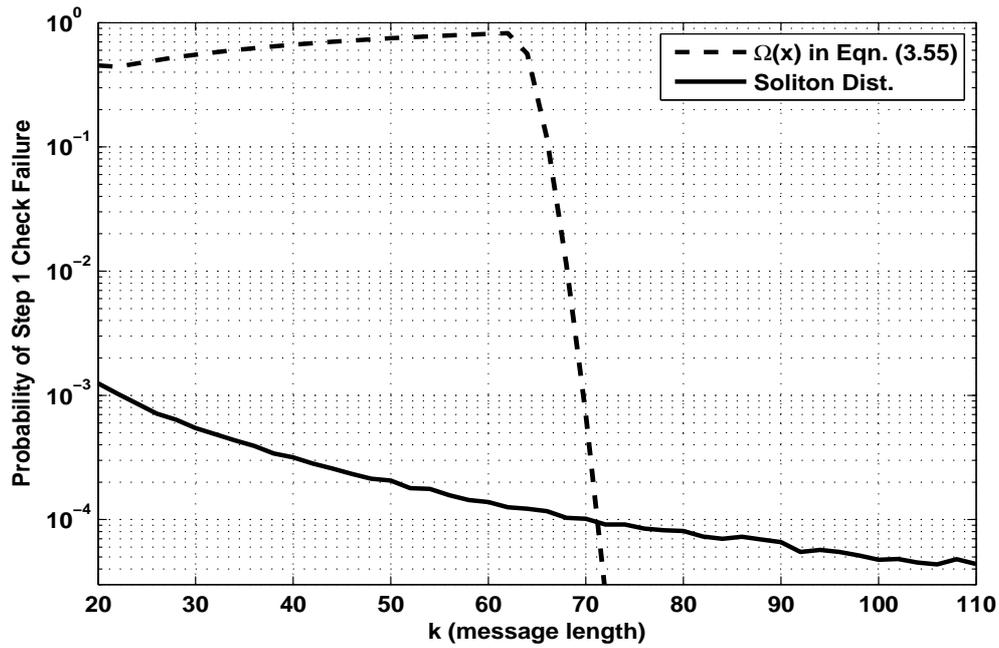}
\caption{Joint probability distribution for order statistics using different degree distributions.}\label{probcomp0}
\end{figure}

\begin{figure}[b!]
\centering
\includegraphics[angle=0, height=35mm, width=135mm]{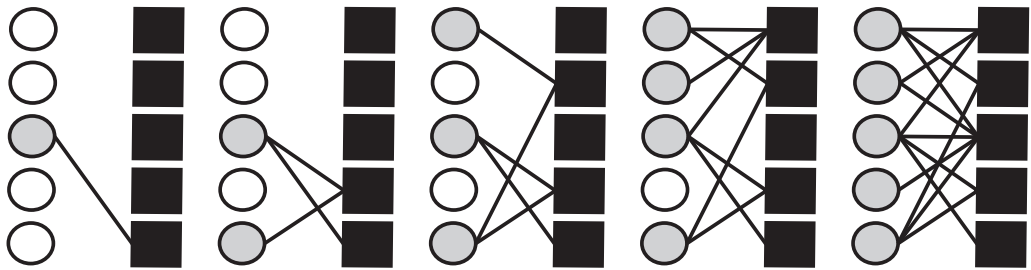}
\caption{An example is shown for $k=5$. At each step which auxiliary block to be recovered is color labeled gray. Connections are made such that in each iteration exactly one auxiliary block is recovered. If there are more than one option for this edge selections, we use uniformly random selections to keep the variable node degree distribution as close to Poisson as possible. }\label{SysLT1}
\end{figure}

\hspace{5mm} We note that if we are unable to check the condition of Step 1 with high probability, we may change the check node degree distribution and it may no longer be close to $\Omega(x)$. To explore this condition let $X_1, X_2, \dots, X_k$ be a sequence of independent and identically distributed $(\sim \Omega(x))$ discrete random variables. Consider the order statistics and define $X_{(i)}$ be the $i$-th minimum of these random variables. Then, we are interested in the joint probability of order statistics that $\Pr\{X_{(3)} \leq 3, X_{(4)} \leq 4, \dots, X_{(k)} \leq k\}$. It can be shown that for check node degree distributions that allow linear encoding/decoding, this probability is close to one as the message length grows. To give the reader a numerical answer, let us compute this probability as a function of message block length $k$ for the check node degree distribution of equation (\ref{RaptorOmega}) as well as for Soliton degree distribution. Note that degree distribution in equation (\ref{RaptorOmega}) allows linear time encoding/decoding whereas Soliton distribution does not. Results are shown in Fig. \ref{probcomp0}. We observe that both distributions allow  a gradual decrease in failure probability as $k$ increases. For example, for $k=1000$ using Soliton distribution the failure rate is around $3 \times 10^{-7}$ although this is not observable in the figure. The degree distribution $\Omega_{*}(x)$ may have to be modified for very large values of $k$ to keep the probability of choosing degree-one non-zero.

\begin{figure}[t!]
\centering
\includegraphics[angle=0, height=38mm, width=118mm]{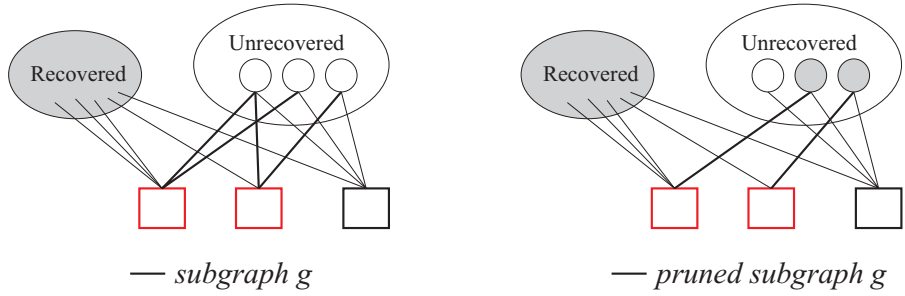}
\caption{Edge eliminations are performed based on the subgraph $g \in G$ each time the BP algorithm gets stuck and need more edge eliminations. }\label{SysFountain1}
\end{figure}

\hspace{5mm} Let us consider an example for $k=5$ and degree vector $\textbf{v} = [1 \ 2 \ 5 \ 2 \ 3]$. It is not hard to show that $\textbf{v}$ can generate an appropriate $G^{*}$. An example of edge selections are shown in Fig.  \ref{SysLT1}. Such a selection is not unique ofcourse. Since some of the selections are not made randomly, this may cause auxiliary node degree distribution little skewed compared to Poisson.

\hspace{5mm} Although the presented approach generates an appropriate $G^*$, it has its own problems with regard to the preservation of node degree distributions. A better approach could be to generate a generator matrix $G$ as in the non-systematic coding, and then eliminate a proper subset of edges so that the BP decoding algorithm can be successful. However, a problem with this approach is that degree distributions of concatenated fountain codes do not allow the full recovery of the message symbols by performing the LT decoding only. Thus, no matter how well we eliminate edges in the graph, the BP decoding will not be successful with high probability unless we use very large overhead $\epsilon$ (which may have dramatic effect on the node degree distributions). An alternative is to insert random edges into the graph $G$ to make each variable node make a connection with the check nodes so that their decoding might be possible. This process corresponds to inserting ones into the appropriate locations in $G$ so that there are no zero rows.

\begin{figure}[t!]
\centering
\includegraphics[angle=0, height=94mm, width=118mm]{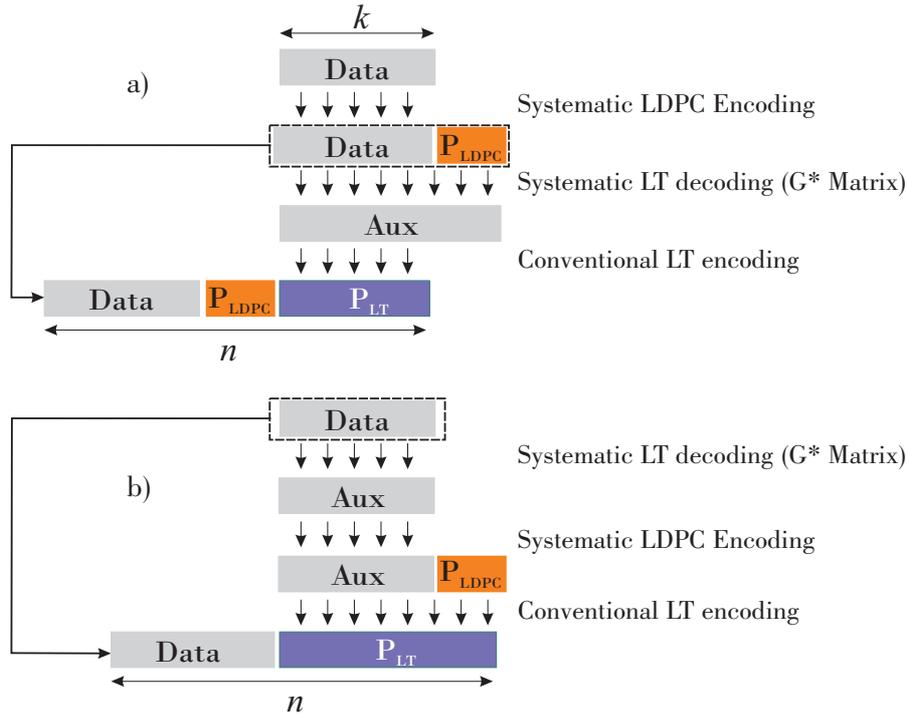}
\caption{Systematic encoding of concatenated fountain codes: An LDPC precode concatenated with an LT base code. }\label{SysFountain2}
\end{figure}

\hspace{5mm} On the other hand, edge eliminations can be performed in various ways \cite{Yuan0}. For example let us start with a full rank $\textbf{G}_{k \times k}$ i.e., $\epsilon = 0$, let $g \in G$ be a subgraph that is induced by the edges connected to unrecovered variable nodes at anytime of the BP algorithm. We eliminate all the edges adjacent to the minimum degree variable node induced by the subgraph $g$ as long as $\textbf{G}$ is full rank. If that edge elimination does not satisfy our condition, the algorithm can check the second minimum degree variable node and so on. This ordered checks guarantee maximum number of unrecovered symbols for BP algorithm to process and decode. A illustrative case is presented in Fig. \ref{SysFountain1} in which the subgraph $g \in G$ is shown with bold edges. As can be seen the rightmost two unrecovered symbols can be recovered after edge eliminations are performed according to the procedure outlined. Sam procedure can be applied to generator matrix with small $\epsilon$ to allow better recovery and performance at the expense of slight change in node degree distributions.

\hspace{5mm} Finally, we show the general guidelines for applying the procedures of this subsection to concatenated fountain codes. Let us consider only two stage coding architecture for simplicity. We can think of two different scenarios of applying $G^*$ to a concatenated fountain code and obtain a systematic construction. The most common architecture is shown in Fig. \ref{SysFountain2} a) where $k$ data symbols are first encoded into $k^{\prime}$ intermediate symbols using systematic LDPC pre-encoding. First $k$ symbols of $k^{\prime}$ intermediate symbols are original data symbols and $k^{\prime}-k$ parity symbols are denoted by $P_{LDPC}$. A $G^*$ is generated according to one of the methods discussed earlier. Using this generator matrix, LDPC codeword is decoded using BP algorithm to form $k^{\prime}$ auxiliary symbols. Note that the BP decoding is successful due to the special structure of $G^*$. Lastly, final parity redundancy is generated as in usual non-systematic LT encoding.  Decoding works in exact reverse direction. First, a large fraction of auxiliary symbols are recovered (not completely! due to linear complexity encoding/decoding constraint) through LT decoding by collecting enough number of encoded symbols. Next, the data and LDPC parity symbols are recovered through LT encoding using only recovered/available auxiliary symbols. Consequently, any remaining unrecovered data symbols are recovered by systematic LDPC decoding.

\hspace{5mm} In Fig. \ref{SysFountain2} b) shows an alternative way of constructing a systematic fountain code. However, it is obvious that this method differ significantly from the former one in terms of decoding performance. It is not hard to realize that in the latter method, data section can only help decode the  auxiliary symbols whereas the parities help decode the whole intermediate symbols, both auxiliary symbols and parities due to LDPC coding. This leads to ineffective recovery of the LDPC codeword symbols. Therefore, the former method can be preferable if the decoding performance is the primary design objective.

\section{\textbf{Advanced Topics}}

\hspace{5mm} In this section, we will discuss some of the advanced topics related to the development of modern fountain codes and their applications in communications and cloud storage systems. We are by no means rigorous in this section as the content details can be found in the references cited. Main objective is rather to give intuition and the system work flow fundamentals. There shall be many more details related to the subject left untouched, which we have not included in this document.

\subsection{Connections to the random graph theory}

\begin{figure}[t!]
\centering
\includegraphics[angle=0, height=45mm, width=79mm]{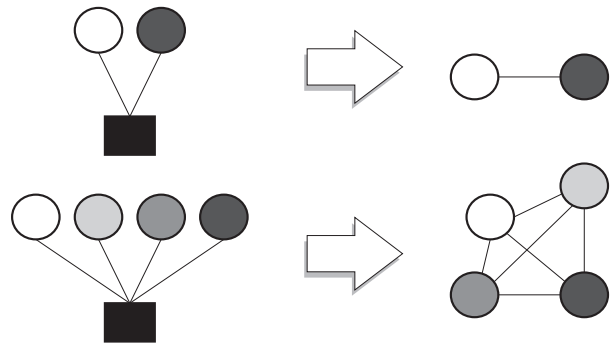}
\caption{Degree-two and degree-four check symbols disappears however the connection properties of the message nodes through these check nodes are preserved. }\label{fig:ERM0}
\end{figure}

\hspace{5mm} The similarity between the random edge generations of a fountain code and the well known random graph models exhibits interesting connections. Before pointing out the similarities, let us state the Erd$\ddot{\textrm{o}}$s-R$\acute{\textrm{e}}$nyi random graph model.

\hspace{5mm} Let $G(k,p_{e})$ be a random graph model such that for  each $\binom{k}{2}$ node pairs, there is an edge connecting these nodes with probability $p_{e}$. All edge connections are performed independently. The degree number of each node is binomially distributed and the average degree per node is $(k-1)p_e$. We are interested in the graphical properties of $G(k,p_{e})$ as $k\rightarrow \infty$. In the original paper \cite{ERmodel}, some of the important asymptotical conclusions drawn about $G(k,p_{e})$  include
\begin{itemize}
\item[$\textbf{j}_\textbf{1}\bullet$] If $kp_e < 1$, then any graph in $G(k,p_{e})$ will almost surely contain connected components of size no larger than $O(\ln(k))$.
\item[$\textbf{j}_\textbf{2}\bullet$] If $kp_e = 1$, then any graph in $G(k,p_{e})$ will almost surely have a largest connected component of size $O(n^{2/3})$.
\item[$\textbf{j}_\textbf{3}\bullet$] If $kp_e = C > 1$ for some constant $C \in \mathbb{R}$, then any graph in $G(k,p_{e})$ will almost surely have a unique giant component containing a fraction of $k$ nodes. The rest of the components are finite and of size no larger than $O(\ln(k))$.
\end{itemize}
\hspace{5mm}  Furthermore, about the connectedness of the whole random graph, they proved for some $\varepsilon > 0$ that
\begin{itemize}
\item[$\textbf{j}_\textbf{4}\bullet$] If $kp_e < (1-\varepsilon)\ln(k)$, then any graph in $G(k,p_{e})$ will almost surely be disconnected.
\item[$\textbf{j}_\textbf{5}\bullet$] If $kp_e > (1+\varepsilon) \ln(k)$, then any graph in $G(k,p_{e})$ will almost surely be connected.
\end{itemize}

\hspace{5mm} Next, we shall show that our conclusions of Section 3.2 and particularly the derivation of Soliton distribution has interesting connections with the set of results in \cite{ERmodel}. First, we realize that the graph representing the fountain code is a bipartite graph. We transform this bipartite graph $G$ to a general graph $G^*$ in which only message nodes exist in the following way. For a degree-$d$ check node, we think of all possible variable node pair connections and thus we draw $\binom{d}{2}$ edge connections between the associated variable nodes (vertices of $G^*_d$ where the subscript specifies the degree-$d$). This procedure is shown for degree-two and degree-four check symbols in Fig. \ref{fig:ERM0}. The transformed graph $G^*$ is thus given by $G^* = \bigcup_d G^*_d$.

\begin{figure}[t!]
\centering
\includegraphics[angle=0, height=31mm, width=115mm]{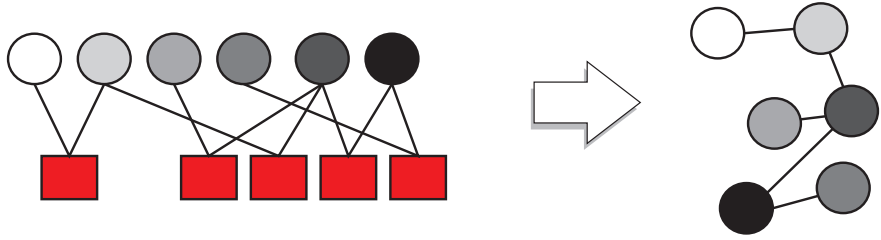}
\caption{A set of degree-two symbols induces a connected graph that can correct every message node if a single degree-one check node is connected to any of the message symbols.}\label{fig:ERM}
\end{figure}

\hspace{5mm} For a given check node degree distribution $\Omega(x) = \sum_{d=1}^k \Omega_d x^d $,  the expected edge connections due to only degree-$d$ check symbols is given by $\binom{d}{2}\Omega_dn$. Since asymptotically the actual number of edge connections converges to the expected value, the probability that an edge exists between two message symbols in $G^*_d$ can be computed to be of the form
\begin{eqnarray}
p_e (d) = \frac{\binom{d}{2}\Omega_dn}{\binom{k}{2}} = \frac{d(d-1)\Omega_d(1+\epsilon)}{k-1}
\end{eqnarray}
if we assume the subgraph induced by using only degree-$d$ check symbols contain no cycles. Of course this assumption might be true for low $d$ such as $d=2$ with high probability. This probability gets lower for large $d$. To see this consider the extreme case $d=k$. Consider the BP algorithm and degree-two check symbols. It is easy to see that the graph $G^*_2$ allows successful decoding of the whole message block if $G^*_2$ is connected and a single degree-one check node is connected to any of the nodes of $G^*_2$\footnote{In fact what node to which this connection is made may change the number of iterations as well as the order of decoded message symbols. The order and the number of iterations of the BP algorithm can be very important in a number of applications \cite{Rahnavard2}, \cite{Sejdinovic}, \cite{Arslan}. }. The requirement of $G^*_2$ being connected is too tight and impractical. We rather impose the connectedness constraint for the overall graph $G^*$. For this to happen, the average degree of a node in Erd$\ddot{\textrm{o}}$s-R$\acute{\textrm{e}}$nyi random graph model is $O(\ln(k))$ (see $\textbf{j}_\textbf{5}$) and hence the average number of edges in the graph $G^*$ must be $O(k\ln(k))$ as established by the arguments of Section 3.

\hspace{5mm} Relaxing connectedness condition, we can alternatively talk about decoding the giant component of $G^*_2$ by making a single degree-one check node connect with any one of the nodes of the giant component of $G^*_2$. In order to have a giant component in $G^*_2$, we must have (according to Erd$\ddot{\textrm{o}}$s-R$\acute{\textrm{e}}$nyi random graph model $\textbf{j}_\textbf{3}$ ),
\begin{eqnarray}
kp_e (2) = \frac{2 k \Omega_2 (1 + \epsilon) }{k-1} > 1 \Longrightarrow \Omega_2 \geq \frac{1}{2}\textrm{ for }k \rightarrow \infty
\end{eqnarray}

\hspace{5mm} Therefore asymptotically, we can decode a fraction $\phi$ of $k$ message symbols if $\Omega_2 \geq \frac{1}{2}$ using only degree-two symbols, where $\phi$ is a function of $\Omega_2$. The exact relationship is found using techniques from random graph theory  \cite{ERmodel2} as characterized by the following theorem,

\hspace{5mm} \textbf{Theorem 8:} \emph{ If $m = kp_e (2) > 1$, the graph  $G^*_2$ almost surely will contain a giant component asymptotically of the size $\phi(m)k$, where
\begin{eqnarray}
\phi(m) = 1 - \frac{1}{m}\sum_{i=1}^{\infty} \frac{i^{i-1}}{i!}(me^{-m})^i.
\end{eqnarray}
}
\hspace{5mm} For convenience, we drop the functional dependence and refer this fraction as $\phi$. Let us consider the remaining $(1-\phi)k$ message nodes and eliminate the edges connected to already decoded $\phi k$ message values. This elimination changes the check node degree distribution and we denote this modified degree distribution by $\Omega_\phi(x)$ from now on. Suppose a check node has originally degree $d$, and its degree is reduced to $i$ after edge eliminations. This conditional probability is simply given by the binomial distribution (since edge selections have been made independently),
\begin{eqnarray}
Pr\{ \textrm{reduced degree} = i | \textrm{original degree} = d  \} = \binom{d}{i}(1-\phi)^i\phi^{d-i}
\end{eqnarray}
for $i=0,1,\dots,d$.  Using standard averaging arguments, we can find the unconditional modified degree distribution to be of the form,
\begin{eqnarray}
\Omega_\phi(x) = \sum_{i=0}^d \sum_{d=1}^k \Omega_d \binom{d}{i}(1-\phi)^i\phi^{d-i} x^i &=&  \sum_{d=1}^k \Omega_d \sum_{i=0}^d  \binom{d}{i}((1-\phi)x)^i\phi^{d-i} \\
&=& \sum_{d=1}^k \Omega_d ((1-\phi)x + \phi)^d \\
&=& \Omega((1-\phi)x + \phi)
\end{eqnarray}

\begin{figure}[t!]
\centering
\includegraphics[angle=0, height=31mm, width=85mm]{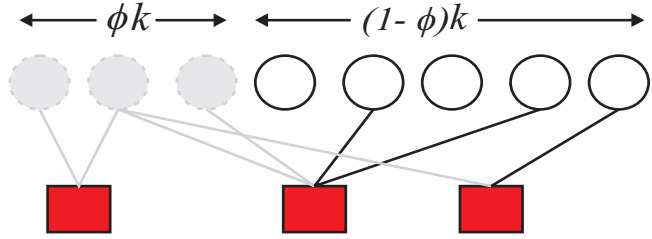}
\caption{A $\phi$ fraction of $k$ message symbols constitute a giant component of the random graph. Elimination of edges of the check nodes that has connections with that giant component induces a modified degree distribution $\Omega_\phi(x)$. }\label{fig:ERM2}
\end{figure}

\hspace{5mm}  In order to find the number of reduced degree-two check nodes, we need to find the probability of having degree-two nodes based on $\Omega_\phi(x)= \Omega((1-\phi)x + \phi)$. Let us look at the Taylor expansion of $\Omega_\phi(x)$ at $x=0$,
\begin{eqnarray}
\sum_{d=0}^{\infty} \frac{\Omega_\phi^{(d)}(0)}{d!}x^d &=&  \Omega_\phi(0) + \frac{\Omega_\phi^{\prime}(0)}{1!}x + \frac{\Omega_\phi^{\prime\prime}(0)}{2!}x^2  + \frac{\Omega_\phi^{(3)}(0)}{3!}x^3 + \dots  \\
&=&\Omega(\phi) +  \frac{(1-\phi)\Omega^{\prime}(\phi)}{1!}x + \frac{(1-\phi)^2\Omega^{\prime\prime}(\phi)}{2!}x^2 + \dots
\end{eqnarray}
from which we find the probability of degree-two to be $\frac{(1-\phi)^2\Omega^{\prime\prime}(\phi)}{2!}$. In order to have a giant component in the remaining $(1-\phi)k$ message nodes, we need to have
\begin{eqnarray}
(1-\phi)kp_e(2) = (1-\phi)k \frac{(1-\phi)^2\Omega^{\prime\prime}(\phi)(1+\epsilon)k }{2! \binom{(1-\phi)k}{2}} =  \frac{ (1-\phi)^2\Omega^{\prime\prime}(\phi)(1+\epsilon)k }{(1-\phi)k-1} > 1
\end{eqnarray}

\hspace{5mm} Finally, some algebra yields
\begin{eqnarray}
(1-\phi)\Omega^{\prime\prime}(\phi) > \frac{1-\frac{1}{k(1-\phi)}}{(1+\epsilon) } \Longrightarrow (1-\phi)\Omega^{\prime\prime}(\phi) \geq 1 \textrm{ for }k \rightarrow \infty \label{5.9}
\end{eqnarray}

\hspace{5mm} First of all in the limit we realize that it is sufficient $\Omega_2 = 1/2$ and $(1-\phi)\Omega^{\prime\prime}(\phi) =1$ to have two giant components and hence an overwhelming portion of the message symbols shall be decoded successfully. Once we set these, we immediately realize that the equation in (\ref{5.9}) is the same as the equation (\ref{KeyEqnOmega}). In fact, we have shown that the degree distribution $\Omega(x)$ satisfying both  $\Omega_2 = 1/2$ and $(1-\phi)\Omega^{\prime\prime}(\phi) =1$ is the limiting distribution of Soliton distribution. In summary therefore, the degree-two check nodes of the Soliton distribution creates a giant component and ensure the recovery of $\phi$ fraction of message symbols whereas the majority of the remaining fraction is recovered by the reduced degree-two check nodes. The decoding to completion is ensured by the higher degree check nodes. This can be seen by applying the same procedure repeatedly few more times although the expressions shall be more complex to track.

\subsection{Inactivation Decoding}

\hspace{5mm} In our previous discussions, we remarked that matrix inversion through Gaussian elimination is costly for optimal fountain code decoding, that is why BP algorithm has become so much popular. However for small $k$, BP decoding algorithm may require large overhead for successful decoding. To remedy this problem, authors in \cite{Amin3} proposed \emph{Inactivation decoding} which combines the ML optimality of the Gaussian elimination with the efficiency of the BP algorithm.

\hspace{5mm} Following their original description, all message symbols are \emph{active} and coded symbols are \emph{unpaired} initially before the BP algorithm is run. The degree of each coded symbols is defined to be the number of active message symbols upon which it depends. At the very start therefore, the degree of each coded symbol is its original degree number. BP is used to find unpaired coded symbol of degree-one and subsequently labels it paired if there is one active message symbol upon which it depends. Executing the update step, paired coded symbol is subtracted from all unpaired coded symbols to which it is connected through the active message symbol. By the end of this process, the degree of the paired coded symbol is reduced by one. BP algorithm repeats this process until either all active message symbols are recovered, or until there is no unpaired coded symbol of degree one, at which point the BP stops. As mentioned earlier, BP algorithm is suboptimal and can fail prematurely although it may still be mathematically possible to decode the whole message block.

\hspace{5mm} In \emph{Inactivation decoding}, when the BP is stuck, the decoder finds one unpaired coded symbol of degree-two if there is any. Then, either one of the active message symbols upon which this coded symbol is connected is declared \emph{inactivated}. Inactivated message symbols are assumed to be known, and therefore the degree of the unpaired coded symbol is now reduced from two to one to allow BP to continue decoding. If there is no degree-two coded symbols, the decoder may search for higher degree coded symbols and deactivate more message symbols. BP algorithm can be stuck after the first inactivation, yet the same inactivation procedure can be applied to make BP continue iterations. If the decoding graph is \emph{ML-decodable}, then this BP algorithm using the inactivation idea will result in successful decoding \cite{Amin3}.

\hspace{5mm} After determining the inactivated $i$ message symbols in the whole process, the $i \times i$ modified submatrix of $\textbf{G}$ is inverted to solve for inactivated symbols since the recovery of unpaired coded symbols depend only on the inactivated message symbols. Eventually, the values for the unrecovered active message symbols can be solved using the belief propagation based on the already decoded message symbols as well as the values of the inactivated message symbols.
This process is guaranteed to recover all of the message symbols if the code is ML-decodable. This modification in the decoding algorithm may call for modifications in the degree distribution design as well. A degree distribution is good if the average degree of edges per coded symbol is constant
and the number of inactivated message symbols are around $O(\sqrt{k})$, as this shall mean that
the total number of symbol operations for inactivation decoding is
linear time complexity.

\hspace{5mm}  Let us give an example to clarify the idea behind inactivation decoding. Let us consider the following system of equations and we would like to decode the message block $\textbf{x}$ from $\textbf{y}$  based on the decoding graph \emph{G} defined by the following generator matrix.

\[ \left( \begin{array}{ccccc}
0 & 0 & 0 & 1 & 1 \\
1 & 0 & 1 & 0 & 0 \\
1 & 1 & 1 & 1 & 0 \\
0 & 0 & 1 & 0 & 0 \\
0 & 1 & 1 & 1 & 1 \\
1 & 1 & 0 & 1 & 1  \end{array} \right) \left( \begin{array}{c}
x_1  \\
x_2 \\
x_3 \\
x_4  \\
x_5   \end{array} \right) =  \left( \begin{array}{c}
y_1  \\
y_2 \\
y_3 \\
y_4  \\
y_5  \\
y_6 \end{array} \right) \]

\begin{figure}[t!]
\centering
\includegraphics[angle=0, height=35mm, width=145mm]{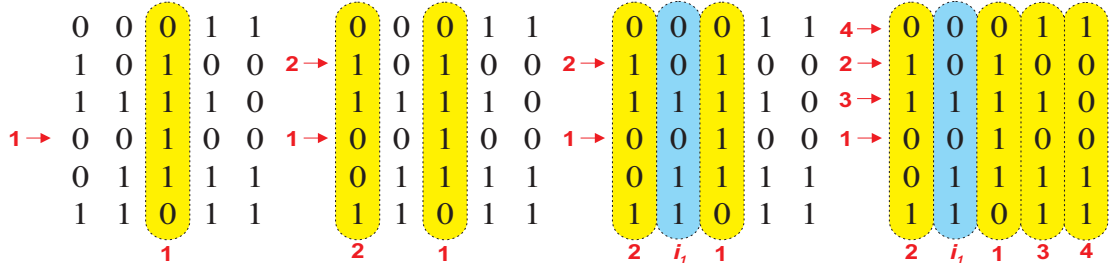}
\caption{BP algorithm worked on the generator matrix of our example. Pink columns show the index of message symbols successfully decoded, i.e., $x_1$ and $x_3$ are recovered. }\label{fig:ID2}
\end{figure}

\begin{figure}[t!]
\centering
\includegraphics[angle=0, height=35mm, width=145mm]{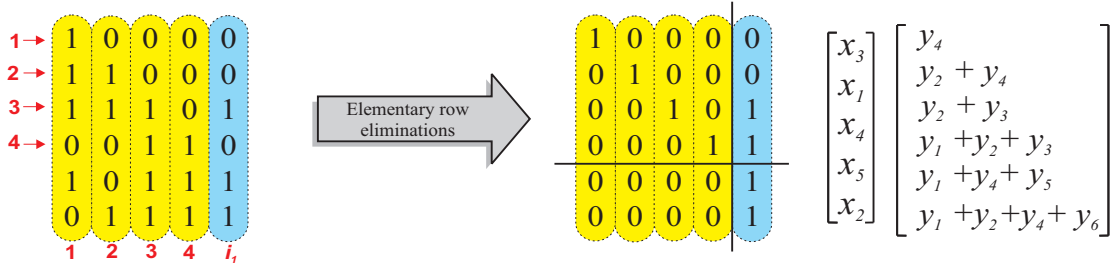}
\caption{BP algorithm worked on the generator matrix of our example. Elementary row operations are performed to obtain the suitable form of the generator matrix where the last entry of the modified $\textbf{x}$ vector is the message symbol that is inactivated. }\label{fig:ID3}
\end{figure}

\hspace{5mm} In the first round of the BP algorithm, only decoder scheduling is performed i.e., what symbols shall be decoded at what instant of the iterative BP algorithm. Let us summarize this scheduling through generator matrix notion. The BP algorithm first finds a degree-one row and takes away the column that shares the one with that row. In order to proceed, there must be at least one degree-one row at each step of the algorithm. This is another way of interpreting the BP algorithm defined in Section 3.2. In our example, BP algorithm executes two steps and gets stuck i.e., two columns are eliminated in the decoding process as shown in Fig. \ref{fig:ID2}. At this point BP algorithm is no longer able to continue because there is no degree-one coded symbol left after edge eliminations (column eliminations). The inactivation decoding inactivates the second column (named $i_1$) and labels it as ``inactive" whereas the previously taken away columns are labeled ``active". Next, we check if BP decoding is able to continue. In our example after inactivation, it continues iterations without getting stuck a second time. Based on the decoding orders of message symbols, we reorder the rows and the columns of the generator matrix according the numbering system shown in Fig. \ref{fig:ID2}. Once we do it, we obtain the matrix shown to the left in Fig. \ref{fig:ID3}. Invoking elementary row operations on the final form of the generator matrix results in a system of equations shown to the right in the same figure. As can be seen, since we only have one inactivated message symbol, the right-bottom of the row echelon form is of size $2 \times 1$. In general if we have $i$ inactivated message symbols and the right-bottom of the row echelon form is of size $n-k+i \times i$. Any invertible $i \times i$ submatrix would be enough to decode the inactivated message symbols. Considering our example,  we have $x_2$ is either given by  $y_1 + y_4 + y_5$ or by  $y_1 + y_2 + y_4 + y_6$. Once we insert the value of $x_2$ into the unrecovered message symbols and run a BP algorithm (this time we allow decoding i.e., XOR operations), we guarantee the successful decoding. In general, the $n-k+i \times i$ matrix is usually dense due to elementary row operations. Hence, the complexity of the inversion operation shall at least be $O(i^2)$. However, row and column operations as well as the elimination steps takes extra effort at least on the order of  $O(i^2)$, thereby making the overall operation at least be $O(i^2)$. That is why if $i \leq O(\sqrt{k})$, linear-time decoding complexity shall still be maintained in the inactivation decoding context.

\hspace{5mm} Alternatively, some of the message symbols can be permanently labeled inactive (and hence the name \emph{permanent inactivation}) before even BP scheduling takes place. Active  and permanently inactive message symbols constitute two class of message symbols and different degree distributions can be used to generate degree and edge connections. This idea in fact resembles to the generalization of LT coding covered in Section 3.6. The usage of permanent inactivations is justified by the observation that the overhead-failure probability curve of the resulting code so constructed is similar to that of a dense random binary fountain code, whereas decoder matrix potentially  has only a small number of dense columns. See \cite{Amin3} for more details.

\subsection{Standardized Fountain Codes for Data Streaming}

\hspace{5mm} The rationale behind the standardized fountain codes for communication applications is to provide the best performance and complexity tradeoff using the most advanced techniques available in the context of fountain codes. These efforts of research community have led to the success of commercialization of some of the concatenated fountain codes covered in previous sections. For example, Raptor 10 (R10) code (an improvement over the first Raptor code in \cite{Amin}) is adapted by 3GPP Multimedia Broadcast Multicast Service, IETF RFC 5053 and IP Datacast (DVB-IPDC) (ETSI TS 102 472 v1.2.1) for DVB-H and DVB-SH. More advanced RaptorQ (RQ) code \cite{Amin3} is implemented and used by Qualcomm  \cite{QualcommRaptor2} in broadcast/multicast file delivery and fast data streaming applications.  Additionally, online codes were used by Amplidata object storage system \cite{Amplidata} to efficiently and reliably store large data sets. We defer the discussion of fountain codes within the context of storage applications for the next subsection.

\hspace{5mm} So far, our discussions were focused on the design of degree distributions and pre-code selections given the simple decoding algorithm, namely BP. Therefore, main objective was to design the code such that the decoding can efficiently be done. We have also seen in Section 3.1 that dense fountain codes under Gaussian elimination provides the best performance if they are provided with a significantly more complex decoding process. Given this potential, the idea behind almost all the advanced standardized fountain codes for communication applications has become to devise methods to get the performance of a concatenated fountain code close to ML decoding performance of a random dense fountain code while maintaining the most attractive feature: low complexity decoding.

\begin{figure}[b!]
\centering
\includegraphics[angle=0, height=45mm, width=135mm]{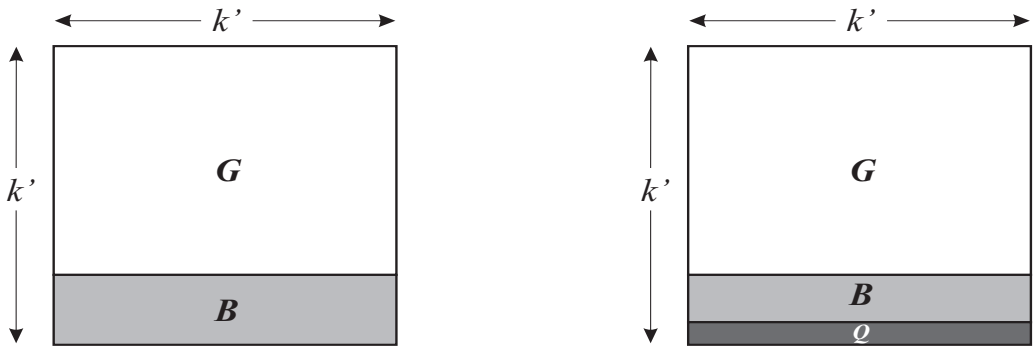}
\caption{Adding dense rows to mimic the failure probability of a random dense fountain
code\cite{Amin3}. Matrices \textbf{G} and \textbf{B} are binary and \textbf{Q} is non-binary. }\label{fig:R10RQ}
\end{figure}

\hspace{5mm} A simple comparison reveals that the code design and degree distributions are functions of the decoding algorithm. R10 and RQ code designs are performed systematically and based on the idea of inactivation decoding. Thus, their designs are little bit different than standard LT and concatenated fountain codes presented so far. One of the observations was that using a small high density submatrix in the sparse generator matrix of the concatenated fountain code successfully mimics the behavior of a dense random linear fountain code. R10 mimics a dense random linear fountain code defined over $\mathbb{F}_2$ whereas RQ mimics a dense random linear fountain code defined over  $\mathbb{F}_{256}$. The generator matrices of these codes have particular structure as shown in Fig. \ref{fig:R10RQ}. As can be seen, a sparse Graph $\textbf{G}$ is complemented by denser matrix $\textbf{B}$ (entries from $\mathbb{F}_2$) in R10 and additionally by $\textbf{Q}$ (entries from $\mathbb{F}_{256}$) in RQ as shown. The design of $\textbf{B}$ is completely deterministic and consist of two submatrices one for sparse LDPC code and one for dense parity check code. These matrices are designed so that BP algorithm with inactivation decoding works well where the majority of design work is mostly based on heuristics. Two potential improvements of RQ over R10 are a steeper overhead-failure curve and a larger number of supported source symbols per encoded source block.  RQ achieves that performance using permanent inactivation decoding and operation over larger field alphabets.

\hspace{5mm} As it has been established in \cite{Liva} and \cite{Schot}, using higher field arithmetic significantly improves the ML decoding performance (also see Section 3.1), thereby making the RQ is one of the most advanced and best fountain codes suitable for data streaming and communication applications. However, although many improvements have been made to make the whole encoding and decoding process linear time for R10 and RQ, the complexity is much higher than concatenated fountain codes defined over binary arithmetic and solely dependent on BP algorithm. Compared to large dense random fountain codes however, R10 and RQ provides significant complexity savings and allows exceptional recovery properties.

\subsection{Fountain Codes for Data Storage}

\hspace{5mm}Erasure codes are used in data storage applications due to massive savings on the number of storage units for a given level of redundancy and reliability. The requirements of erasure code design for data storage, particularly for distributed storage applications, might be quite different relative to communication or data streaming scenarios. One of the requirements of coding for data storage systems is the systematic form, i.e., the original message symbols are part of the coded symbols. As covered in Section 3.5, systematic form enables reading off the message symbols without decoding from a storage unit. In addition to this property, since bandwidth and communication load between storage nodes can be a bottleneck, codes that allow minimum communication could be very beneficial particularly when a node or a symbol fails and a system reconstruction is initiated. This is usually referred as the \emph{locality} of the erasure code in literature \cite{locality}.

\hspace{5mm} For a largely scalable system, fountain codes and their on-the-fly symbol generation (rateless) capability can make them one of the most popular choices for distributed storage applications. However, with the aforementioned  requirements, the fountain code design will have to change dramatically. Particularly in connection with the locality requirements, fountain codes are expected to allow efficient repair process in case of a failure: when a single encoded symbols is lost, it should not require too much communication and computation among other encoded symbols in order to resurrect that lost symbol. In fact, repair latency might sometimes be more important than storage space savings. However, the code overhead $\epsilon$ and the repair efficiency have conflicting goals and hence an optimal code with respect to code overhead is frequently not optimal with respect to repair efficiency. To see this, let us start with the following definition.

\hspace{5mm} \textbf{Definition 5:} (\emph{Repair Efficiency}) Let $\mathcal{L}$ be a subset of the fountain code symbols of size $(1+\epsilon)k$ to be repaired. The repair complexity is defined to be the average number of symbol operations per repaired symbol by the repair algorithm. In otherwords, it is the total number of symbol operations in repairing $\mathcal{L}$ divided by $|\mathcal{L}|$.

\hspace{5mm} Let us consider the repair efficiency of the previously introduced fountain codes. A straightforward repair algorithm will require the decoding of the whole message block from any $(1+\epsilon)k$ coded symbols. Upon successful decoding, missing symbols in  $\mathcal{L}$ can be generated through fountain code encoding. Let $\mu_e$ and $\mu_d$ be the encoding the decoding cost per symbol, respectively. The decoding stage thus requires $\mu_dk$ average symbol operations. The encoding symbols in $\mathcal{L}$ takes us an average of $\mu_e|\mathcal{L}|$ symbol operations. The repair complexity is given by,
\begin{eqnarray}
C_R(\mathcal{L}) = \frac{\mu_dk + \mu_e|\mathcal{L}|}{|\mathcal{L}|} = \frac{\mu_dk }{|\mathcal{L}|} + \mu_e
\end{eqnarray}
which might have been efficient if $|\mathcal{L}|$ was on the order of $k$. However, in many practical cases single failure or two are more frequent. Even if $\mu_d$ and $\mu_e$ are constant, the repair process will be inefficient for large $k$ and small constant $|\mathcal{L}|$ i.e., $C_R(\mathcal{L}) = O(k)$.

\hspace{5mm} One can immediately realize that the main reason for inefficient repair is tied to the fact that the message symbols (auxiliary symbols in the systematic form) are not readily available when a failure is attempted to be corrected and thus the decoding of the whole message block (or the whole auxiliary block in the systematic form) is necessary. To remedy this in \cite{Gummadi}, a copy of the message block is stored in addition to non-systematic fountain code symbols. In this case, it is shown that the expected repair complexity for an arbitrary set $|\mathcal{L}|$ is at most $(1+\epsilon)\Omega^{\prime}(1)$. For Raptor codes for instance $C_R(\mathcal{L}) = O(1)$. Therefore, a constant repair complexity can be achieved. However, the overall cost is the increased overhead $\epsilon^{\prime} = 1+\epsilon$, which does not vanish as $k \rightarrow \infty$.

\hspace{5mm} In \cite{Asteris}, the existence of fountain codes have been shown that are systematic and has a vanishing overhead and a repair complexity $C_R(\mathcal{L}) = O(\ln(k))$. These codes are defined over $\mathbb{F}_q$ and the construction is based on the rank properties of the generator matrix. Since the BP algorithm is not applicable to the encoding and decoding, the complexity of the original encoding and decoding processes is high. The existence of systematic fountain codes, preferably defined over $\mathbb{F}_2$, with very good overhead and repair complexity, while accepting a low complexity encoding/decoding is an open problem.


%
%
%
%

\newpage
\begin{thebibliography}{100} 
\bibitem{Shannon0} C. E, Shannon, ``A mathematical theory of communication," \emph{Bell Sys. Tech. J.,} vol. 27,
pp. 379-423; July 1948, pp. 623-625, Oct. 1948.
\bibitem{MacKay} D. J. C. MacKay, ``Fountain codes," \emph{in IEE Proceedings Communications,} vol.
152, no. 6, 2005, pp. 1062--1068.
\bibitem{Spinal} J. Perry, P. Iannucci, K. Fleming, H. Balakrishnan, and D. Shah. ``Spinal
Codes," \emph{In SIGCOMM}, Aug. 2012.
\bibitem{Hagenauer}  J. Hagenauer, ``Rate-Compatible Punctured Convolutional
Codes (RCPC Codes) and Their Applications," \emph{IEEE Trans.
Commun.}, vol. 36, no. 4,  pp. 389--400, Apr. 1988.
\bibitem{RCPT}  A. S. Barbulescu and S. S. Pietrobon, ``Rate compatible turbo codes,"
\emph{Electron. Let.,} pp. 535–536, 1995.
\bibitem{RCLDPC} J. Ha, J. Kim and S. W. McLaughlin, ``Rate-compatible puncturing of
low-density parity-check codes," \emph{IEEE Trans. Inform. Theory,} vol. 50,
no. 11, pp. 2824–2836, Nov. 2004.
\bibitem{Oberg} M. Oberg, P. H. Siegel, ``The Effect of Puncturing in Turbo Encoders",
\emph{in Proceedings of the International Symposium on Turbo Codes and Related
Topics,} pp. 204--207, Brest, France, 1997.
\bibitem{Byers} J. W. Byers, M. Luby, M. Mitzenmacher, A. Rege, ``A Digital Fountain Approach to Reliable Distribution of Bulk Data", \emph{In Proc. of  SIGCOMM}, pp. 56--67, 1998.
\bibitem{Luby} M. Luby, ``LT codes," \emph{in Proc. 29th Annu. ACM Symp.
Theory of Computing},  pp. 150--159, 2002.
\bibitem{Di} C. Di, D. Proietti, E. Telatar, T. Richardson, and R. Urbanke, ``Finite Length Analysis
of Low-Density Parity-Check Codes," \emph{IEEE Trans. on Information Theory,} pp. 1570--
1579, Jun. 2002.
\bibitem{Liva}G. Liva, E. Paolini, and M. Chiani, ``Performance versus Overhead for
Fountain Codes over $\mathbb{F}_q$," \emph{IEEE Communications Letters,} vol. 14, no. 2,
pp. 178--180, 2010.
\bibitem{Rahnavard} N. Rahnavard and F. Fekri, ``Bounds on Maximum-Likelihood Decoding
of Finite-Length Rateless Codes," \emph{In Proc. of the 39th Annual Conference
on Information Science and Systems (CISS'05)}, March 2005.
\bibitem{Schot} B. Schotsch, R. Lupoaie, and P. Vary, ``The Performance of Low-Density
Random Linear Fountain Codes over Higher order Galois fields under
Maximum Likelihood Decoding," \emph{In Proc. 49th Annual Allerton Conf. on
Commun., Control, and Computing,} Monticello, IL, USA,
pp. 1004--1011, Sep. 2011.
\bibitem{Petar} P. Maymounkov, ``Online Codes," \emph{Secure Computer Systems Group,} New York Univ., New York, Tech. Rep. TR2002--833, 2002.
\bibitem{Luby2} M. Luby, M. Mitzenmacher, and A. Shokrollahi, ``Analysis of Random Processes via And-Or
Tree Evaluation", \emph{In Symposium on Discrete Algorithms}, 1998.
\bibitem{Oliver} Oliver C. Ibe, ``Elements of Random Walk and Diffusion Processes", Wiley Series in Operations Research and Management Science, 2013.
\bibitem{Amin} A. Shokrollahi, ``Raptor Codes," \emph{IEEE Trans. Inf. Theory,} vol. 52, no. 6, pp. 2410--2423, Jun. 2006.
\bibitem{Dimakis} A. G. Dimakis, P. B. Godfrey, Y. Wu, M. J. Wainwright and K. Ramchandran `` Network coding for distributed storage systems," \emph{IEEE Trans. Inf. Theor.,} 56(9):4539--4551, Sept. 2010.
\bibitem{Yuan0} X. Yuan and L. Ping, ``On systematic LT codes," \emph{IEEE Commun. Lett.,}
vol. 12, pp. 681--683, Sept. 2008.
\bibitem{Rahnavard2} N. Rahnavard, B. N. Vellambi, and F. Fekri, ``Rateless codes with
unequal protection property," \emph{IEEE Trans. Inf. Theory,} vol. 53, no. 4,
pp. 1521--1532, Apr. 2007.
\bibitem{Sejdinovic} D. Sejdinovic, D. Vukobratovic, A. Doufexi, V. Senk, and R. Piechocki,
``Expanding window fountain codes for unequal error protection," \emph{IEEE
Trans. Commun.}, vol. 57, no. 9, pp. 2510--2516, Sep. 2007.
\bibitem{Arslan} S. S. Arslan, P. C. Cosman, and L. B. Milstein, ``Generalized unequal error
protection LT codes for progressive data transmission," \emph{IEEE Trans.
Image Processing}, vol. 21, no. 8, pp. 3586--3597, Aug. 2012.
\bibitem{Arslan2} S. S. Arslan, P. C. Cosman, and L. B. Milstein, ``Optimization of Generalized LT Codes for Progressive Image Transfer,"   \emph{IEEE Visual Communications and Image Processing (VCIP)}, pp. 1--6, San Diego, Nov. 2012.
\bibitem{Ron} Ron M. Roth, ``Introduction to Coding theory", Cambridge Univ. Press., 2006.
\bibitem{QualcommRaptor} Qualcomm Incorporated White Paper, `` Why Raptor Codes Are Better Than Reed-Solomon Codes for Streaming Applications", Jul. 2013.
\bibitem{Luby3} M. Luby, M. Mitzenmacher, A. Shokrollahi, D. Spielman , V. Stemann, ``Practical Loss-Resilient Codes", \emph{In Proc. of the twenty-ninth annual ACM symposium on Theory of computing (STOC):} pp. 150--159, 1997.
\bibitem{Bloemer} J. Bloemer, M. Kalfane, M. Karpinski, R. Karp, M. Luby and D.
Zuckerman, ``An XOR-Based Erasure-Resilient Coding Scheme",
ICSI Technical Report, TR-95-048, August 1995.
\bibitem{Amin2} A. Shokrollahi, ``LDPC Codes: An Introduction," 2003. \emph{Available online:} http://www.awinn.ece.vt.edu/twiki/pub/Main/AthanasNotebook/LDPC\_Codes\_-\_An\_Introduction.pdf
\bibitem{Johnson} S. J. Johnson, ``A Finite-Length Algorithm for LDPC Codes Without Repeated Edges on the Binary Erasure Channel", \emph{IEEE Transactions on Information Theory}, vol. 55, no. 1, pp. 27--32, 2009.
\bibitem{Zyablov}  V. Zyablov, R. Johannesson,   M. Loncar, and  P. Rybin, ``On the Erasure-Correcting Capabilities of Low-Complexity Decoded LDPC Codes with Constituent Hamming Codes," \emph{In Proc. 11th Int. Workshop on Algebraic and Combinatorial Coding Theory (ACCT'2008),} Pamporovo, Bulgaria,  pp. 338--347, 2008.
\bibitem{Jin} H. Jin, A. Khandekar, and R. McEliece, ``Irregular Repeat-Accumulate
Codes," \emph{In Proc. 2nd Int. Symp. Turbo Codes,} Brest, France, pp. 1--8, Sep. 2000.
\bibitem{Yuan} X. Yuan and L. Ping, ``Doped Accumulate LT Codes," \emph{In Proc. IEEE Int.
Symp. Inf. Theory (ISIT) 2007}, Nice, France, pp. 2001--2005, Jun. 2007.
 \bibitem{ERmodel}  P. Erd$\ddot{\textrm{o}}$s and A. R$\acute{\textrm{e}}$nyi, ``On the Evolution of Random Graphs", \emph{Publications of the Mathematical Institute of the Hungarian Academy of Sciences} \textbf{5}: 17--61, 1960.
\bibitem{ERmodel2}  P. Erd$\ddot{\textrm{o}}$s and A. R$\acute{\textrm{e}}$nyi, ``On Random Graphs", Publicationes Mathematicae 6 pp. 290--297, 1959.
\bibitem{Amin3} A. Shokrollahi and M. Luby, ``Raptor Codes", \emph{Foundations and
Trends in Communications and Information Theory,} vol. 6, no.
3-4, pp. 213--322, 2009.
\bibitem{QualcommRaptor2} Qualcomm Incorporated White Paper, `` RaptorQ Technical Overview ", 2010. \emph{Available online:}
http://www.qualcomm.com/solutions/multimedia/media-delivery/raptorq
\bibitem{Amplidata} K. D. Spiegeleer and  R. R. A. Slootmaekers, ``Method of Storing a Data Set in a Distributed Storage System, Sistributed Storage System and Computer Program Product for Use with Said Method", U.S. Patent 2011/0113282 A1, May 12, 2011.
\bibitem{locality} P. Gopalan, C. Huang, H. Simitci, and S. Yekhanin, ``On the Locality of Codeword Symbols," \emph{IEEE Transactions on Information Theory,}
vol. 58, no. 11, pp. 6925--6934, 2012.
\bibitem{Gummadi} R. Gummadi and R. S. Sreenivas, ``Erasure Codes with Efficient Repair," 2012.  \emph{Available online:}
http://www.stanford.edu/~gummadi/papers/repair.pdf
\bibitem{Asteris} M. Asteris and A. G. Dimakis, ``Repairable Fountain Codes," \emph{In
Proc. of 2012 IEEE International Symposium on Information Theory,}
Cambridge, MA, July 2012.
\end{thebibliography}
\end{document}